**The Artificial Personality Model: Theory, Implementation, Measurement and Experimental Validation in Conversational Agents**

**Alexander Dregger[1,2]*, Barbarella Petz[3], Erik Kuhn[4], Pia Volk[1], Andreas Oberweis[1,2], Stefan Hillmann[5], Jannik Strötgen[4]**

**Affiliations**
[1] Institute of Applied Informatics and Formal Description Methods (AIFB), Karlsruhe Institute of Technology (KIT), Kaiserstr. 89, 76133 Karlsruhe, Germany
[2] Innovation, Strategy and Transfer, FZI Research Center for Information Technology, Haid-und-Neu-Str. 10–14, 76131 Karlsruhe, Germany
[3] Department of Computer Science, Humboldt-Universität zu Berlin, Rudower Chaussee 25, 12489 Berlin, Germany
[4] Faculty of Computer Science and Business Information Systems, Karlsruhe University of Applied Sciences, Moltkestr. 30, 76133 Karlsruhe, Germany
[5] Quality & Usability Lab, Technische Universität Berlin, Marchstr. 23, 10587 Berlin, Germany
***Corresponding author:** Alexander Dregger, Email: alexanderdregger@googlemail.com

**Abstract**

Artificial personality shapes user experience (UX) in human–machine interaction, yet coherent theoretical frameworks and validated measurement instruments remain limited. This paper refines and empirically evaluates the Artificial Personality Model (APM), a hybrid framework integrating human-centred and machine-specific personality dimensions. Across three experiments (N = 395), the model was operationalised and evaluated using the Artificial Personality Questionnaire (APQ). Factor analyses revealed that some human personality dimensions (e.g., agreeableness, conscientiousness) are partially transferable, whereas others (e.g., extraversion) did not emerge as distinct dimensions. Additional machine-specific dimensions were identified. The APQ provided encouraging initial psychometric evidence, while the human-centred BFI-2 showed poor reliability and an uninterpretable factor structure in Experiment 3. Experimental manipulations demonstrated significant effects of artificial personality on user satisfaction, continuous use intention, and social presence. The findings contribute an empirically refined framework and a novel instrument for measuring perceived artificial personality, offering practical guidance for designing conversational agents.

## 1. Introduction

Over 50% of the 8,000 users of robotic vacuum cleaners surveyed by the company Shark Robot give their vacuum cleaner a name, and 63% of them claim that their vacuum cleaner has a personality (Hopwood, 2025). Although the study is not representative, it underlines that machines are more than mere tools for humans: they are social actors according to the Computers are Social Actors (CASA) paradigm (Nass et al., 1994). Humans anthropomorphise such systems and regard them as social beings, as this facilitates interaction with machines and enables users to adapt their behaviour accordingly (Epley et al., 2008).

Artificial personality (AP) can influence key user experience outcomes, including trust (Zhou et al., 2019), acceptance (Tapus et al., 2008), and purchase decisions (Shumanov & Johnson, 2021). Beyond end users, AP is increasingly considered relevant in the design of CAs, with practitioners highlighting its potential role in creating consistent agent behaviour, adapting interactions to user needs, fostering long-term user–system relationships, and guiding design decisions. Despite this relevance, fundamental questions remain unresolved, in particular: how can artificial personality be defined, implemented, and measured?

In general, there are two types of models to describe the artificial personality of conversational agents (Dregger, 2023): human-centred and machine-centred models. Human-centred models transfer personality models developed for humans to the context of machines. In research and practice, human-centred models are dominant. The Big Five Model (BFM), also called the Five-Factor Model (FFM) of personality (Costa & McCrae, 2008), is one of the most frequently applied approaches for designing linguistic cues for CAs (Ahmad et al., 2022; Dregger et al., 2024), and widely applied in studies on artificial personality (Mou et al., 2020; Völkel & Kaya, 2021). Besides the BFM, the Sixteen Personality Factor Questionnaire (Cattell & P. Cattell, 1995) is also mentioned (Endres, 1995). Both models follow a trait-based approach to personality.

However, alternative approaches also exist in psychology. The psychoanalytic paradigm and the transactional paradigm are likewise used to describe artificial personality. Psychoanalytic approaches (e.g. Myers-Briggs-Inventory, Myers & Myers, 1995) are used in practice but lack robust empirical validation. In contrast, the transactional paradigm conceptualizes personality as an interaction between individual and environment. Thus, they consider not only personality traits, but also situational variables. This perspective is reflected in approaches such as character computing (Herbert, 2020; Hundertmark et al., 2023). However, its higher complexity limits its direct applicability in design contexts, which may explain the dominance of trait-based models (Dregger, 2023).

Machine-centred models question whether "human personality models can be transferred to machines" (Dregger, 2023, p. 4; Völkel et al., 2020). One example is the model by Völkel et al. (2020), which was specifically developed for conversational agents using experiments and analysis of customer reviews. It describes ten dimensions of personality: dysfunctional, confrontational, unstable, serviceable, approachable, social-entertaining, social-inclined, social-assisting, self-conscious, and artificial. Another approach was developed by Kovačević, Holz, et al. (2024) that encompasses eight dimensions. More recently, Contreras (2026) created an LLM-native psychometric instrument using a bottom-up approach and derives five factors from LLM responses: responsiveness, deference, boldness, guardedness, and verbosity.

Both human-centred and machine-centred approaches face several criticisms:

- **Human personality traits have a limited validity for machines:** Not all human personality dimensions are meaningful in machine contexts. Experts consider, for example, neuroticism (emotional stability) and openness to experience as less relevant for the design of the AP of conversational agents (Dregger, 2023). Importantly, this represents a construct-level problem that is distinct from the question of whether individual questionnaire items can be reformulated for machines. Recent research demonstrates this distinction: Zierahn et al. (2026) found that LLM-adapted Big Five inventories, such as the BFI-LLM (Tosato et al., 2025) and LMLPA (Zheng et al., 2025), can achieve high content validity at the item level, while the theoretically assumed five-factor structure nevertheless could not be recovered empirically. Thus, adapting the operationalisation of human personality constructs does not necessarily establish that the underlying dimensions themselves constitute an appropriate model of artificial personality.
- **The BFM provides a limited perspective on personality and consequently, is a problematic foundation for the development of a personality model for machines:** The BFM focusses on five personality dimensions, and these may not capture human personality in a fully valid manner. Paunonen & Jackson (2000) argue that at least ten additional dimensions can be identified based on the same data. Accordingly, Paunonen (2002) developed a supernumerary personality model including ten additional dimensions. In the expert study by Dregger (2023), at least two of these dimensions, namely humour and honesty, were considered highly relevant for machines but are not included in the BFM.
- **The transfer of human-centred models into a design of a system is difficult:** Human-centred models are often considered too abstract to derive concrete design decisions for conversational agents (CAs). Machine-centred models such as Völkel et al. (2020) may address this limitation by considering machine-specific characteristics; however, they also provide limited guidance for implementation. An exception is the model by Kovačević, Boschung, et al. (2024), which gives explicit recommendations for prompting personality.
- **The machine-centred models overlap considerably with the human-centred models:** Machine-centred approaches, such as the one by Völkel et al. (2020), substantially overlap with the dimensions of the BFM (Dregger, 2023). For example, the "agreeableness" dimension in the BFM is reflected in multiple dimensions in Völkel et al. (2020): approachable, social-inclined, and social-assisting. This reduces parsimony and complicates the transfer of empirical findings from BFM-based studies to machine-centred models (Dregger, 2023). Similar overlaps are observed in Kovačević, Holz, et al. (2024), where dimensions such as profundity or instability overlap with conscientiousness and emotional stability. The authors themselves appear to acknowledge this overlap, as the model was reduced from eight to five dimensions in later work (Kovačević, Boschung, et al., 2024). Contreras (2026) derives five LLM-native factors—responsiveness, deference, boldness, guardedness, and verbosity—without imposing dimensions from human personality theory. While such an approach avoids assuming that human personality structures directly transfer to machines, empirically emerging factors are not necessarily theoretically novel personality constructs. Machine-centred dimensions should therefore be systematically compared with established personality constructs before conceptual distinctiveness is assumed. Otherwise, bottom-up approaches risk rediscovering established personality characteristics under new labels.

- **Measurement is a big challenge in both approaches due to several reasons:** Artificial personality is often designed but not systematically measured (Shumanov & Johnson, 2021; Smestad, 2018). A key reason is the lack of validated and reliable measurement instruments. For machine-centred models, user-based measurement instruments are largely missing. Existing approaches were mainly used for theory development and rely on adjective-based scales that provide limited guidance for design improvement (Kovačević, Boschung, et al., 2024; Völkel et al., 2020). For human-centred models, studies (Holtgraves et al., 2007; Ruane et al., 2021; Völkel & Kaya, 2021) often use human personality questionnaires such as the BFI-2 (Soto & John, 2017). Although the BFI-2 was developed to measure human personality, authors do not report how they adapt the questionnaire to the context of machines. Furthermore, essential quality criteria such as reliability or factor structure are not reported (Ruane et al., 2021; Völkel & Kaya, 2021). In addition, many BFI-2 items were rated as not valid by experts (Dregger, 2023). Recently, the BFI was adapted for LLMs with specific questionnaires, such as the BFI-LLM (Tosato et al., 2025) and the LMLPA (Zheng et al., 2025). Zierahn et al. (2026) compared established human Big Five inventories with these new inventories. While the original BFI-44 (John et al., 1991) and IPIP-NEO-120 (Johnson, 2014) showed insufficient content validity for LLMs, two adapted instruments achieved high content validity. However, subsequent factor analysis still failed to recover the expected Big Five structure. This distinction is crucial: items may constitute plausible descriptions of machine behaviour without the dimensions to which these items are assigned forming an empirically valid personality structure for machines. Consequently, the transferability of human personality models needs to be examined at two levels: the applicability of their measurement items and, more fundamentally, the applicability of the personality dimensions themselves. The APM addresses both levels by not assuming a priori that the complete dimensional structure of a human personality model transfers to artificial systems.

Another problem arises from the lack of proper measurement instruments: artificial personality can be implemented via language, voice, movement, facial expression, haptics, or proxemics in conversational agents (Dregger et al., 2024; Mou et al., 2020). However, without valid measurement methods, findings on these implementation strategies remain uncertain. The same applies to the use of large language models and prompting techniques (Bhattacharjee et al., 2024; Jiang et al., 2022; Kovačević, Boschung, et al., 2024).

Considering these limitations of both human-centred and machine-centred approaches, Dregger (2023) introduces a third type of personality model for conversational agents and machines in general, namely the Artificial Personality Model (APM), which can be understood as a hybrid approach combining both perspectives as seen in **Table *1***.

**Table 1**

*Typology of Personality Models for the Description of Artificial Personality*

| Human-centred Models | Machine-centred Models | Hybrid Models |
|---|---|---|
| Big Five (Costa & McCrae, 2008; Soto & John, 2017) and its adaptations for LLMs BFM-LLM (Tosato et al., 2025) and LMLPA (Zheng et al., 2025)<br>Archetypes<br>MBTI (Myers & Myers, 1995) | Völkel et al. (2020)<br>Kovačević, Holz, et al. (2024)<br>Contreras (2026) | Artificial Personality Model (Dregger, 2023) |

| | |
|---|---|
| Character Computing (Herbert, 2020; Hundertmark et al., 2023) | |

Due to the shortcomings of the human-centred and machine-centred models, this paper focusses on the Artificial Personality Model as the model could lead to more valid results regarding the description, implementation and measurement of artificial personality. Thus, the paper aims to achieve four main goals:

**Research Goal 1: Conceptualisation:** The APM consists of seven dimensions (extraversion, agreeableness, conscientiousness, humour, honesty, artificiality, and serviceability), but only three of them are currently described in terms of facets. These facets are essential for enhancing model precision and facilitating practical application when designing a CA.

**Research Goal 2: Measurement:** Given the lack of valid and reliable measurement instruments and the resulting questionable empirical findings this paper aims to develop a questionnaire for measuring artificial personality.

**Research Goal 3: Developing an LLM-based approach for the implementation of the APM in CAs:** To evaluate the model, it can be implemented in empirical studies using either mock-up systems or live systems. The implementation in this paper uses both a mock-up system and two LLM-based live systems. The different implementation strategies will be tested, and recommendations for the implementation will be derived.

**Research Goal 4: Experimental Evaluation:** To assess the model and the questionnaire, three experiments will be designed. Those experiments focus on different domains and different dimensions of the APM. This is necessary to show that the model is robust in different contexts. In addition, current experiments suffer from several methodological issues: Besides not using a valid measurement instrument, the experiments (Ruane et al., 2021; Völkel & Kaya, 2021) do not have a control group and hence the impact of the manipulation of the experiment on the user experience is unclear. This paper suggests how to solve this issue.

This paper is structured as follows. First, the theoretical background is presented, covering relevant concepts and theories such as anthropomorphism, the CASA paradigm, artificial personality, conversational agents (CAs), and approaches to implementing artificial personality in CAs. Subsequently, the methodological approach is outlined, with particular emphasis on the development of the questionnaire. The experimental procedures and findings are then presented in the results section. Following this, the results of the three experiments are discussed, and the revision of the model and questionnaire is described. The paper then elaborates on the theoretical contributions and practical implications of the findings while also addressing limitations and avenues for future research. Finally, the conclusion summarizes the key findings and contributions of the study.

## 2. Background and Related Work

Several theoretical concepts are required in light of the research goals of this paper. First, foundational concepts underlying artificial personality are introduced, namely anthropomorphism and the Computers are Social Actors (CASA) paradigm. Second, artificial personality is defined. Third, theories and methods related to the implementation of artificial personality are discussed, including conversational agents, language cues, mock-ups, and large language models (LLMs). Finally, the chapter addresses how artificial personality can be measured.

### 2.1. Anthropomorphism and Computers are Social Actors paradigm (CASA)

The term anthropomorphism was first described by the Greek philosopher Xenophanes (Xénophanes & Lesher, 2001). It can be defined as the "attribution of distinctively human-like feelings, mental states, and behavioural characteristics to inanimate objects, animals, and, in general, to natural phenomena and supernatural entities" (Salles et al., 2020). Anthropomorphism can be observed in both users and developers of AI systems (Salles et al., 2020).

The basis for this attribution lies in the human ability to perceive physical features and behaviours beyond what is directly observable. According to Epley et al. (2008), two motivations foster this tendency:

- **Sociality motivation:** Humans are inherently social beings. In the absence of other humans, non-human entities, objects, and natural phenomena are anthropomorphised to fulfil the need for social interaction.

- **Effectance motivation:** Humans need to understand and control their environment (White, 1959) and by anthropomorphising their environment they can reduce their perceived uncertainty regarding the behaviour of a non-human entity by projecting human qualities on it. By projecting these, humans can make assumptions about the behaviour and use those assumptions to control and understand the environment.

Humans differ in their tendency to anthropomorphise, and factors such as age and personality play a role (Letheren et al., 2016). The anthropomorphic response itself depends not only on such individual traits but also on macro-environmental factors such as culture and system design (Kim & Im, 2023). Furthermore, the design of the system, including the artificial personality, plays an important role in triggering anthropomorphism.

In line with the concept of sociality motivation proposed by Epley et al. (2008), the "Computers are Social Actors" (CASA) paradigm assumes that computers can be perceived as social actors (Nass et al., 1994). According to CASA, humans exhibit social responses towards machines similar to those shown in human–human interaction. One key factor triggering such responses is the use of language by the machine (Fogg, 2003).

The paradigm was developed in the 1990s, its core assumption remains valid (Xue et al., 2026), although limitations may exist, for example in the context of desktop PCs (Heyselaar, 2023). Furthermore, contextual factors must be considered (Kaptelinin & Dalli, 2025). Extensions of the model have been proposed (Gambino et al., 2020; Lombard & Xu, 2021), helping to explain why the paradigm remains relevant even with modern technologies such as social robots.

### 2.2. Artificial Personality and the Artificial Personality Model

One of the key components of anthropomorphism and the CASA paradigm is personality, as users attribute personality to machines (Holtgraves et al., 2007; Nass et al., 1994, 1995). Thus, it is necessary to develop a theory that enables the systematic design of artificial personality as users attribute it anyway to a system. Several terms exist to describe this concept, such as chatbot personality (Ahmad et al., 2022) or robot personality (Mou et al., 2020). In contrast, artificial personality (AP) is more suitable as a general term, as it is independent of specific technologies and describes the overarching concept (Dregger, 2023). It can be defined as "the relatively stable tendencies and patterns of behaviour of a (AI-based) machine that can be designed by developers and designers via different modalities, such as language, creating the impression of individuality of a humanized social agent when users interact with the machine." (Dregger, 2023)

Artificial personality encompasses four qualities (Dregger, 2023):

- **Focus on behaviour:** Human personality describes the relatively stable tendencies and patterns regarding the behaviour and the experience of persons (Rauthmann, 2017). In the context of artificial personality only the behaviour of the machine is considered. This is the case because the artificial personality theory does not assume that machines have consciousness. There is a huge debate in the literature on the issue of consciousness in machines (Krauss & Maier, 2020).
- **Designed by humans:** Artificial personality can be intentionally implemented by developers and designers. However, even if it is not explicitly designed, humans still perceive personality. Nevertheless, to what extent it actually can be designed and how stable this design is has to be further examined.
- **Individuality of a humanized social agent:** Due to the various possibilities in personality design, machines can give users the impression of individuality, even while fulfilling the same task.
- **Perception impacts the UX:** The perception of artificial personality influences users' behaviour, cognition, and attitudes, thereby affecting the overall user experience.

Based on an expert study, Dregger (2023) introduced the artificial personality model. It integrates dimensions of the human-centred personality model BFM and the supernumerary personality indices (SPI) and of machine-centred models like the model proposed by Völkel et al. (2020). It consists of seven dimensions, as it can be seen in Figure 1: extraversion, conscientiousness, agreeableness, humour, honesty, artificiality, and serviceability. Only for the dimensions extraversion, conscientiousness, and agreeableness, facets were defined as those dimensions are based on the BFM. Nevertheless, the model has not been fully conceptualised as facets for four dimensions and for all dimensions, definitions and a questionnaire to measure them are missing. Additionally, the model has not been empirically tested.

**Figure 1**

*Original Artificial Personality Model without Facets for the Dimensions Humour, Honesty, Artificiality, and Serviceability*

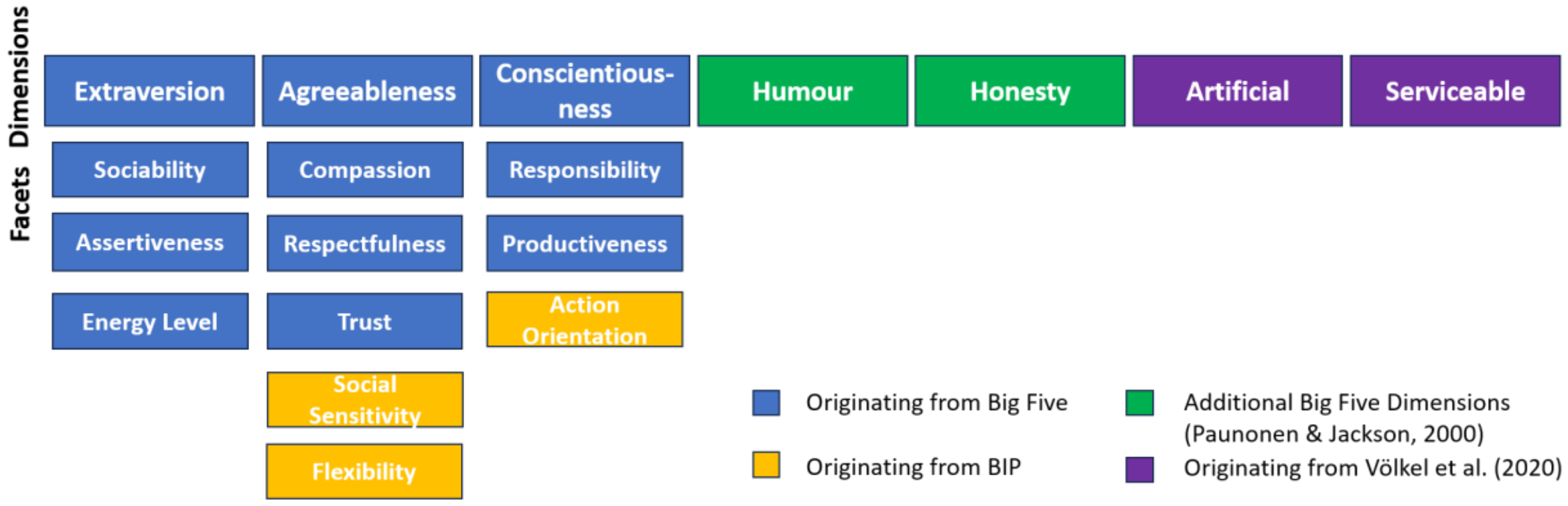

*Note 1 Reproduced from Dregger (2023)*

Artificial personality and human personality can interact in two ways:

- **Similarity attraction hypothesis** (also called similarity attraction or law of attraction): Originating from social psychology (Byrne, 1997), the hypothesis assumes that similar personalities not only between humans, but also between a human and a machine attract each other and facilitate the interaction (Nass et al., 1995; Nass & Lee, 2001)
- The **complementary hypothesis** (also called complementary attraction or convergence effect) has a different approach: Humans prefer to interact with a machine that has a complementary personality to their own (Isbister & Nass, 2000). If, for example, a human has a low conscientiousness, the human prefers a machine with a high conscientiousness.

Empirical evidence supports both hypotheses (for the similarity hypothesis: Nass et al., 1995; Nass & Lee, 2001; for the complementary hypothesis: Isbister & Nass, 2000; Jin & Youn, 2021). Both may hold true, as an important moderating factor is often not considered: social roles (Joosse et al., 2013). Machines fulfil different social roles, such as salesperson, consultant, or friend. Depending on these roles, users may prefer either similar or complementary personalities (Joosse et al., 2013).

Apart from similarity, the attraction between a machine and a human is also influenced by other factors such as proximity (mere exposure effect), physical attractiveness (the beautiful is good effect), and liking (Edwards et al., 2022).

### 2.3. Conversational Agents

According to chatbot.org (Walker, 2026), 161 synonyms exist for the term conversational agent, ranging from chatbot to social robot to interactive voice response (IVR). In general, conversational agents are “providing access to information and services through interaction in everyday language” (Følstad et al., 2021). The variety of terminology can be explained by the different types of CAs, which can be distinguished based on various criteria (Adamopoulou & Moussiades, 2020), such as embodiment, capabilities, goal and the technology to generate and process the dialogue.

This paper focuses on text-based chatbots with virtual embodiment. One of the systems is implemented as a mock-up, while the other two use LLM-based response generation and all three systems focus on a specific domain. The domains are very diverse to test the robustness of the theory.

### 2.4. Implementation of Artificial Personality into CAs using social cues, mock-ups and LLMs

**Figure 2**

*Process of the Perceiving Artificial Personality in the Interaction between a Conversational Agent and a User*

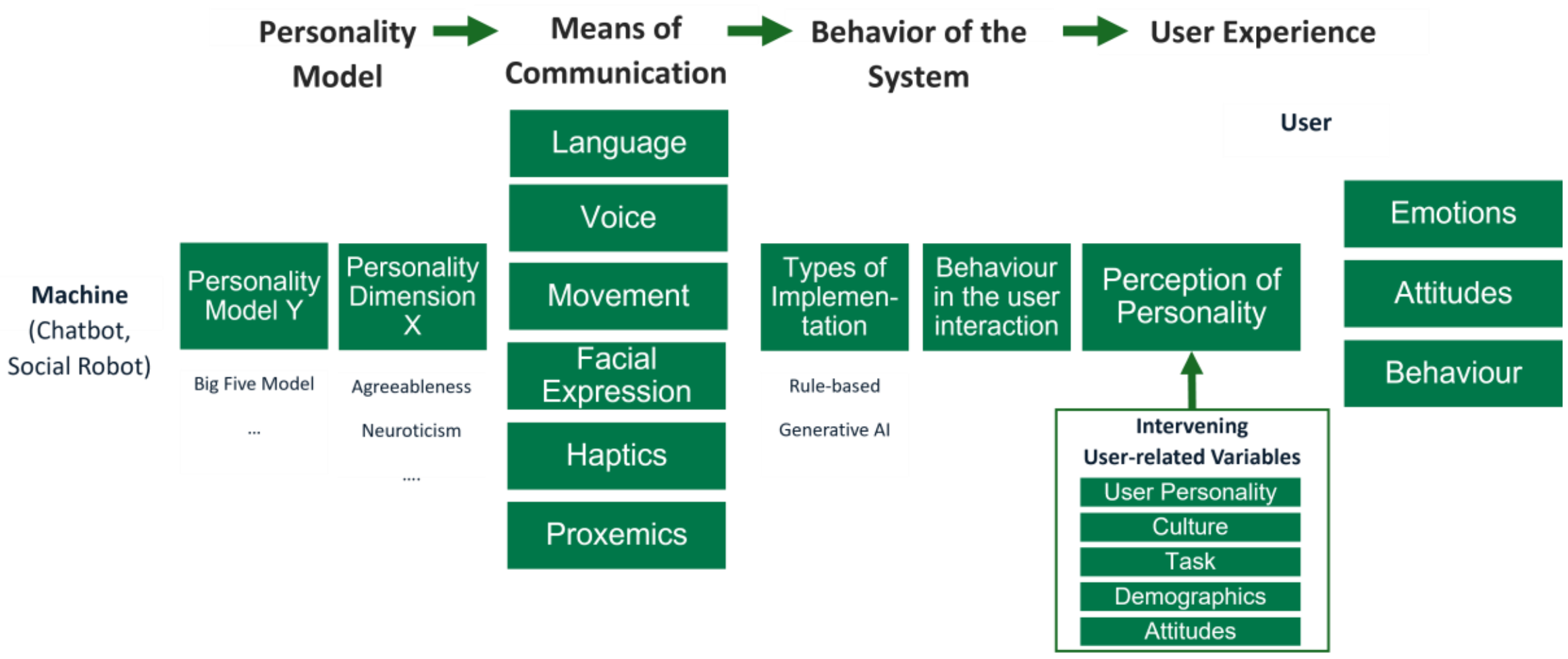

Artificial personality can be expressed through different modes of communication, such as language, voice, or proxemics (Mou et al., 2020), as illustrated in Figure *2*. However, not all modalities can be considered in this paper as this would increase the complexity and limit the precision of the analysis. Thus, this paper focusses on language as it is a very central form of expression of human communication, and it is the core of the chatbots which will be the technology for implementation in this paper. With regard to language, different types of social cues play a role, including verbal, visual, auditory, and invisible cues (Feine et al., 2019), as illustrated in Figure 3.

As this paper concentrates on chatbots, verbal, visual, and invisible cues are particularly relevant for the expression of artificial personality. Consequently, chatbot responses within a conversation constitute the primary medium

**Figure 3**

*Overview of Social Cues based on Feine et al. (2019)*

| Verbal Cues | | | Visual Cues | | | | Auditory Cues | | Invisible Cues | |
|---|---|---|---|---|---|---|---|---|---|---|
| Content | Content | Style | Kinesics | Proxemics | Appearance | CMC | Voice Qualities | Vocalizations | Chronemics | Haptics |
| Apology | Self-disclosure | Abbreviations | Arm and hand gesture | Background | 2D-/3D-agent visualization | Buttons | Pitch Range | Grunts and moans | First turn | Tactile touch |
| Asking for permission | Self-focused questions | Dialect | Eye movement | Conversational distance | Age | Emoticons | Voice gender | Laughing | Reponse time | Temperature |
| Greeting and farewell | Small talk | Formality | Facial movement | | Attractiveness | Typeface | Voice tempo | Vocal segregates | | |
| Joke | Tailored responses | Lexical diversity | Head movement | | Clothing | Typos | Volume | Yawn | | |
| Name | Thanking | Response variety | Posture shift | | Color of Agent | Typing Indicator | | | | |
| Opinion conformity | Tips and advice | Sentence complexity | | | Degree of human-likeness | | | | | |
| Praise | | Strength of Language | | | Facial Feature | | | | | |
| Referring to Past | | | | | Gender | | | | | |
| | | | | | Name Tag | | | | | |
| | | | | | Photorealism | | | | | |

*Note 2 Reproduced from Dregger et al. (2024), https://doi.org/10.1145/3640794.3665559*

through which personality is expressed. These responses can be generated using different technological approaches, which in turn affect the possibilities for implementing artificial personality:

- **Mock-ups/Clickable prototypes:** Artificial personality can be implemented through clickable mock-ups, which are interactive conceptual visualisations of a system at the GUI level (Böhm & Graser, 2023). While low-fidelity mock-ups provide a general representation, high-fidelity mock-ups allow detailed dialogue flows in which personality can be conveyed through verbal, visual, and invisible cues (Dregger et al., 2024). Although mock-ups provide a standardised setting, their predefined interactions restrict user interaction and may reduce immersion and the experience of artificial personality.
- **Retrieval-based systems:** Retrieval-based systems respond to user inputs relying on predefined rules and a database of responses (Adamopoulou & Moussiades, 2020). Artificial personality can be conveyed through these predefined responses but requires knowledge of how specific social cues relate to personality dimensions and facets. Moreover, limited flexibility in responding to unforeseen inputs constrains both interaction and personality expression.
- **Generative systems:** Large language models (LLMs) provide a promising foundation for generating artificial personality in conversational agents (Ait Baha et al., 2023). Three techniques can be used to implement artificial personality via LLMs:
  - **Training of a language model:** LLMs can be trained on personality-labelled datasets to learn associations between personality traits and language cues. However, this requires high-quality datasets and substantial financial and technical resources (Ait Baha et al., 2023).
  - **Fine-tuning:** Existing foundation models can be adapted using personality-labelled datasets. Although less resource-intensive than full training, fine-tuning remains costly (Ait Baha et al., 2023).
  - **Prompt Engineering:** Personality can be adapted through text-based prompts without modifying the model itself (Marvin et al., 2024). While resource-efficient, its outcomes can be difficult to predict (Marvin et al., 2024).

This paper uses prompt engineering as the most resource-efficient approach. Specifically, zero-shot personality prompting (Jiang et al., 2022) and personality infusion (Kovačević, Boschung, et al., 2024) are applied. Both use in-context learning, in which model behaviour is conditioned through information provided in the prompt rather than changes to model parameters (Brown et al., 2020). Alternative approaches are excluded because they either do not rely on a personality model (Sun et al., 2024) or do not use LLMs (Ramirez et al., 2023).

### Zero-shot Personality Prompting

Zero-shot prompting refers to prompting a language model without providing example inputs and outputs (Brown et al., 2020). Instead, the desired behaviour is specified solely through instructions given in the prompt. Jiang et al. (2022) use GPT3.5 and their prompting technique is based on the big five personality model (Costa & McCrae, 2008). The prompt consists of three elements:

- **Naïve prompt:** This component specifies the desired personality trait based on the Big Five model (e.g. “you are extroverted”).
- **Keyword Prompt:** Descriptive terms derived from psychological research are used to characterise the selected trait (e.g. “You are talkative, outgoing, energetic, enthusiastic, social…”).
- **Personality Prompt:** Based on the keywords, the LLM generates short descriptive statements representing individuals with the specified personality trait (e.g. “You are a very friendly and outgoing person who loves to be around others. You are always…”).

In addition to the personality description, the prompt includes contextual information and a task or question for the model.

### Personality Infusion

The personality infusion method developed by Kovačević, Boschung, et al. (2024) is based on a personality model introduced by Kovačević, Holz, et al. (2024). This model comprises eight dimensions: decency, profoundness, instability, vibrancy, engagement, neuroticism, serviceability, and subservience. In the study evaluating the personality infusion method, only five dimensions are used: vibrancy, conscientiousness, civility, artificiality, and neuroticism (Kovačević, Holz, et al., 2024). In contrast to other prompting approaches, personality infusion is implemented across multiple LLMs, including GPT-3.5, Llama 2 13B, and Mistral 7B. Furthermore, it separates the generation of personality from the generation of content. The prompt consists of four elements, namely context (description of the personality model), task (what is the role of the agent?), output format and additional data (includes the conversational context and dialogue history).

In contrast to many other prompting techniques, the personality infusion was validated with actual user testing.

### 2.5. Measurement of Artificial Personality

Two approaches for measurement of AP can be identified in the literature (Ait Baha et al., 2023; Jiang et al., 2022):

- **Rating by the machine itself / automatic evaluation:**
    - **Self-evaluation by the machine using a machine-adapted questionnaire**: Based on the Big Five Model (Costa & McCrae, 2008) and the IPIP dataset (Goldberg et al., 2006), Jiang et al. (2022) developed the Machine Personality Inventory (MPI). In this approach, the machine itself rates the items of the MPI. Similar to that, Zierahn et al. (2026) let several LLMs evaluate LLM-adapted versions of the BFM. This method is problematic for several reasons. First, the user experience of personality can only be validly assessed when humans are involved, not when the machine evaluates itself, as machines only represent humans and this representation is always limited and can be biased. Recent empirical evidence provides further support for this concern. Contreras (2026) developed a bottom-up, LLM-native psychometric instrument specifically to avoid the potential mismatch caused by transferring human personality dimensions to LLMs. Although the resulting five-factor self-report structure demonstrated high internal consistency and replicability, self-report scores generally failed to predict corresponding behaviour as evaluated by human observers. Thus, the discrepancy between machine self-evaluation and externally perceived behaviour cannot be explained solely by the use of personality constructs developed for humans. Even machine-centred constructs may produce internally consistent self-reports without adequately representing how a system actually behaves and is perceived by users. Second, the Big Five Model (BFM) and the IPIP items were developed for humans; therefore, applying them to machines constitutes a problematic approach to describing and measuring artificial personality (Dregger, 2023), particularly as certain dimensions, such as neuroticism, are irrelevant in the context of human–machine interaction. Furthermore, artificial personality differs from human personality and includes dimensions, such as artificiality, that are not part of the IPIP. This approach implicitly assumes that human and artificial personality are identical, an assumption that can be questioned (Dregger, 2023; Kovačević, Boschung, et al., 2024; Völkel et al., 2020). Zierahn et al. (2026) support these claims as they compared empirically five Big Five-based inventories regarding their content validity for LLMs. The MPI achieved an S-CVI of only .29 for chatbot suitability, identical to the IPIP-NEO-120 and substantially below the commonly applied threshold of .80. In contrast, more extensively adapted Big Five inventories achieved substantially higher content validity. Thus, describing an instrument as machine- or LLM-oriented does not by itself establish its validity for artificial systems. More fundamentally, successful item adaptation does not establish the validity of the underlying personality model. Although the LLM-adapted inventories examined by Zierahn et al. (2026) achieved high content validity, the expected five-factor structure could not subsequently be confirmed. Therefore, two conceptually distinct validation questions need to be considered: whether individual items appropriately describe machine behaviour and whether the theoretically assumed personality dimensions adequately structure this behaviour. Third, human-centred personality models provide limited guidance for translating personality traits into concrete design decisions for conversational agents. Finally, Jiang et al. (2022) compare the machine's scores with average values reported by Johnson (2014) and interpret these as representative of how "humans" behave. However, Johnson (2014) did not use a representative sample of humans, which makes such generalisations problematic.
    - **Estimation of personality based on classifying dialogue text:** This approach uses linguistic features extracted from dialogue and applies classifiers to map these features onto personality dimensions (Mairesse et al., 2007). This method is also problematic, as the machine effectively

evaluates itself rather than being evaluated by human users. Consequently, it remains unclear whether such evaluations reflect human perception of artificial personality.

- **Ratings by humans / human evaluation**: In this approach, the personality of the machine is evaluated either using personality items from human personality questionnaires or so-called personality descriptors. These descriptors typically consist of adjectives associated with personality dimensions. This method is employed, for example, by Kovačević, Holz, et al. (2024) and Völkel et al. (2020). However, it comes with several limitations: First, the personality descriptors used in these studies are based on machine-centred personality models developed by the respective authors. Machine-centred models are frequently constructed using bottom-up approaches. While this allows machine-specific behavioural patterns to emerge without imposing human personality structures, it also carries a risk of limited theoretical integration. Empirically identified factors are not necessarily novel personality constructs merely because they emerge from machine data. Existing approaches may partly rediscover characteristics already captured by established personality theories. For instance, the dimensions "approachable", "social-inclined", and "social-assisting" identified by Völkel et al. (2020) are conceptually close to agreeableness. More recent LLM-native approaches likewise derive new factors from machine data (Contreras, 2026), highlighting the need to systematically examine whether newly labelled dimensions are conceptually distinct from established personality constructs. Second, the use of personality descriptors in the form of isolated adjectives is problematic because these models define their dimensions primarily through lists of descriptors rather than through theoretically grounded constructs. As a result, the meaning of the dimensions remains ambiguous. Moreover, practitioners require actionable insights for the design of conversational agents. Without behaviour-based items and clearly defined theoretical constructs underlying the dimensions and their facets, the implications for design remain vague and difficult to operationalise.

Overall, these theoretical perspectives provide the basis for understanding how conversational agents can convey human-like characteristics. While anthropomorphism and the CASA paradigm emphasize human-like cues in human–technology interaction, artificial personality focuses specifically on conveying individuality and personality. According to the APM, AP is expressed exclusively through behavioural characteristics and aims to create an impression of individuality that influences the user experience. AP can resemble or complement the user's personality and can be conveyed through different technological implementations. This paper examines how AP can be conveyed through language in LLM-based systems and mock-ups, focusing on verbal, visual, and invisible cues. For LLM-based implementation, personality infusion and zero-shot personality prompting are applied. To investigate the effects of AP design, a new questionnaire is developed to measure perceived artificial personality. As no established instrument specifically captures AP in conversational agents, its development represents an important methodological step and is described in the following chapter.

## 3. Methods

This paper focuses on the relationship between artificial personality and user experience when interacting with a CA. To analyse this, the Artificial Personality Model developed by Dregger (2023) serves as the foundation. However, the model is not yet fully developed, as descriptions of facets are missing for four dimensions, as shown in Figure 1. Furthermore, the dimensions have not been clearly defined, no questionnaire exists to measure the model, and no implementation methods have been described. These shortcomings are addressed in this paper.

As a first step, the model is further specified by developing and defining facets for the dimensions that currently lack them. This development is based on a literature review. In a second step, a questionnaire is developed by identifying items from the literature, while also adding newly formulated items. The first version of the questionnaire is then evaluated by experts. Based on this evaluation, the questionnaire is refined.

After developing both the theoretical framework and the questionnaire, they are tested in three experiments. This approach is chosen for three reasons. First, the model is examined across different domains and samples to assess the extent to which its structure generalises across contexts. Second, the model is tested using different implementation methods: the first experiment relies on zero-shot personality prompting (Jiang et al., 2022), the second on a mock-up, and the third on personality infusion (Kovačević, Boschung, et al., 2024). Third, the aim is to examine the influence of the APM on UX. To analyse different aspects of UX, the experiments focus on different user experience variables.

Chapter 3 will focus on the theory and questionnaire development and Chapter 4 will present the methods of the experiments and the results.

## 3.1. Development of a Questionnaire

The APM is based on the dispositional or trait-based paradigm, which is one of the seven paradigms of personality psychology (the others being psychodynamic, learning-theory-based, humanistic, cognitive, biological, and transactional). This dispositional paradigm assumes that personality can be described in terms of traits (Rauthmann, 2017). These traits, also referred to as personality dimensions, capture specific aspects of personality but are relatively broad. Therefore, they can be further specified through facets (Rauthmann, 2017).

The APM consists of seven personality dimensions. Three dimensions originate from the Big Five Model (BFM) (Soto & John, 2017), two from the supernumerary personality indices (SPI) (Paunonen, 2002), and two from the machine-centred personality model of Völkel et al. (2020). Some facets from the BIP (Hossiep & Paschen, 2003) were also included (see Figure 1).

For each dimension, the theoretical background is outlined, and facets are identified based on a literature review. Subsequently, an overview of the revised model is provided, forming the basis for the questionnaire development.

### 3.1.1. Extraversion

Extraversion is not only a key component of the BFM, but also of other dimensional personality theories such as HEXACO (Ashton et al., 2004), 16PF (Cattell & P. Cattell, 1995), and PEN (Eysenck, 1998). In humans, it is defined as “a preference for companionship and social stimulation“ (McCrae & Costa, 1999, p. 143). In the context of machines, it was rated together with the dimension “agreeableness” as the most relevant dimension by experts in the study by Dregger (2023) and is also the most frequently studied personality dimension regarding implementation in CAs (Ahmad et al., 2022; Dregger et al., 2024).

Several facets are described in the literature (Rauthmann, 2017), including assertiveness, activity, warmth, gregariousness, positive emotions/energy, excitement-seeking, social self-esteem, social boldness, sociability, and liveliness. Ten facets would be too many for efficient measurement, and some overlap conceptually. Therefore, the number is reduced to three:

- **Sociability** is considered highly relevant in the context of machines, as it indicates whether users perceive a system as open to social interaction.
- **Assertiveness** reflects whether a system behaves dominantly in an interaction, which is important as excessive assertiveness may negatively affect the interaction.
- **Positive Energy** captures the perceived energy level of the system; higher levels may enhance interaction quality.

### 3.1.2. Agreeableness

This dimension is part of both the BFM (McCrae & Costa, 1999) and the HEXACO model (Ashton et al., 2004). In humans, it is defined as “primarily a dimension of interpersonal behaviour. Highly agreeable individuals are trusting, sympathetic and cooperative” (Costa & McCrae, 1992, p. 6).

In human–computer interaction, it is the third most studied dimension in conversational agents regarding implementation via language cues (Dregger et al., 2024) and together with “extraversion” the most relevant dimension according to experts (Dregger, 2023).

Numerous facets have been identified (Ashton et al., 2004; Rauthmann, 2017), including trust, altruism, tender-mindedness, straightforwardness, compliance, modesty, forgiveness, gentleness, flexibility, and patience. To ensure parsimony, this list is reduced to five facets, supplemented by social sensitivity from the BIP:

- **Compassion** was regarded by the experts as very important (Dregger, 2023), and reflects whether the system conveys empathy and concern for user well-being.
- **Respect** was also viewed as very important by the experts (Dregger, 2023) and indicates whether the system treats users considerately and avoids antagonistic behaviour.

- **Trust** was seen in the study with the experts by Dregger (2023) as important and reflects whether the system appears non-malicious and does not assume malicious intent from the user.
- **Social sensitivity** is originally a facet related to the Bochum Inventory for Job-Related Personality Description (BIP) model (Hossiep & Paschen, 2003) that was studied in the expert study of Dregger (2023) and this facet was evaluated as very relevant by the experts. It is conceptually very close to agreeableness and thus it was attributed to this dimension. It describes the system's ability to respond appropriately to "social" dynamics during interaction.
- **Flexibility** is a part of the BIP, but also of the HEXACO model and in the study of Dregger (2023) it was regarded as the most relevant facet. It reflects the system's ability to adapt to changes in interaction.

#### 3.1.3. Conscientiousness

Conscientiousness is part of both the BFM (McCrae & Costa, 1999) and the HEXACO model (Ashton et al., 2004). In humans, it "contrasts scrupulous, well-organised, and diligent people with lax, disorganised, and lackadaisical individuals" (Costa & McCrae, 1992, p. 6). It is the second most studied personality dimension in terms of verbal cue implementation (Dregger et al., 2024) and the fourth most relevant dimension in Dregger (2023). Facets identified in the literature include competence, dutifulness, achievement striving, self-discipline, deliberation, orderliness, organisation, diligence, perfectionism, and prudence (Ashton et al., 2004; Rauthmann, 2017). Due to overlap and measurement efficiency, the following facets are selected:

- **Responsibility:** The facet responsibility was rated highly by the experts in the study of Dregger (2023) and reflects whether the system reliably fulfils obligations and follows instructions.
- **Productiveness:** It describes persistence and goal-oriented behaviour of the system.
- **Action orientation:** It is part of the BIP model and reflects how efficiently the system implements user decisions.

#### 3.1.4. Humour

Humour is not part of the BFM, but Paunonen & Jackson (2000) reanalysed datasets that were used for the development of the BFM and found 10 additional dimensions that might not be covered by the BFM, but could be important to describe personality (Paunonen & Jackson, 2000; Rauthmann, 2017). Based on these dimensions, Paunonen (2002) developed the supernumerary personality indices (SPI) and an accompanying questionnaire. In the SPI humour is defined as "an individual's ability to arouse amusement and laughter in people. It also involves the capacity for recognizing and reacting to something amusing or funny in others" (Paunonen, 2002, p. 7).

Experts rated it among the most important dimensions for describing artificial personality in Dregger (2023), and recent research has increasingly examined humour in conversational agents (Brîncoceanu, 2025; Zargham et al., 2023). According to Brîncoceanu (2025), humour must be personalised for enhancing engagement and has to consider context, ethical boundaries and the effects on trust. Zargham et al. (2023) recognise a complexity on a technical level, but they demand a more intensive study of humour when using AI to create enjoyable interactions with the machines. When humour is analysed in experimental studies in CAs, they sometimes lack a theoretical conceptualisation. A good example for that is the study of Bhattacharjee et al. (2024). The study does not reflect that there are different types of humour (Martin et al., 2003) in humans and such types of humour could also play an important role in the design of humour in machines. Without such a differentiation it is difficult to properly prompt an LLM in a valid manner and thus, the validity of the results can be questioned.

In contrast to the BFM related dimensions, no facets have been described regarding the personality dimension humour in machines. Hence, a literature review was done to identify potential facets based on theoretical models and concepts of human humour. The following theoretical aspects of humour have been described in the literature:

Human research conceptualises humour both as a relatively stable individual characteristic and as observable behaviour (Ruch, 2010). For the APM, only aspects that can be expressed through observable machine behaviour are relevant. Two components are therefore distinguished: the tendency to recognise and express humour and the style in which humour is expressed.

The first component is represented by **sense of humour**. Existing conceptualisations describe sense of humour as a multidimensional construct that includes recognising, appreciating, and expressing humorous content (McGhee, 1999; Svebak, 2010). To maintain parsimony and avoid overlap with the behavioural humour styles, the APM captures this construct more narrowly as the machine's general tendency to recognise humorous potential in an interaction and respond accordingly.

The second component concerns humour style. Martin et al. (2003) distinguish affiliative, aggressive, self-enhancing, and self-defeating humour. Their framework is used because it is well established and differentiates humour according to its interpersonal function. Three of these styles can be transferred to artificial personality:

- **Affiliative Humour** describes humorous behaviour intended to facilitate social interaction and reduce tension.
- **Aggressive Humour** refers to sarcasm, teasing, or other humorous behaviour expressed without sufficient consideration of its potential negative impact on the user.
- **Self-defeating Humour** describes humorous behaviour in which the system makes itself the object of jokes in order to amuse the user or facilitate the interaction.

Self-enhancing humour is excluded because it primarily serves an intrapersonal function by regulating an individual's own emotional state in stressful situations (Martin et al., 2003). Such an internally oriented function is inconsistent with the APM's focus on observable machine behaviour and does not require the assumption that an artificial system possesses subjective emotional experiences.

Accordingly, the humour dimension of the APM comprises four facets: **sense of humour, affiliative humour, aggressive humour, and Self-Defeating humour**. This conceptualisation goes beyond treating artificial humour as a single characteristic and allows different forms of humorous machine behaviour to be designed and evaluated separately.

Humour competence, which could be regarded as another facet, is deliberately excluded from the APM. The ability to determine whether a joke is appropriate or to generate high-quality humour may be relevant for human–machine interaction, but it describes an aptitude rather than a relatively stable behavioural tendency (Guilford, 1959). The APM focuses on temperament-like behavioural tendencies; humour competence should therefore be considered separately from artificial personality.

#### 3.1.5. Honesty

Honesty can be conceptualised both as behaviour and as a trait (Fleeson et al., 2022). It is included in models such as SPI (Paunonen, 2002), VIA (McGrath, 2019), and HEXACO (Ashton et al., 2004). It refers to the "inhibition of stealing, cheating, and deceiving behaviors" even if this is connected with a potential detriment to oneself (Paunonen, 2002, p. 7). It is also connected to have integrity and authenticity and telling the truth (McGrath, 2019). Paunonen (2002) uses the term "integrity" for the dimension instead of honesty.

There is a lack of clearly defined facets in the literature especially in the context of machines. Thus, this paper develops facets to describe honesty in the context of machines:

- **Openness of intentions**: The system explains which goals it is pursuing, and which functions it has proactively or on demand.
- **Transparency**: The system clarifies proactively or on request where and how it processes data.
- **Compliance**: The system clearly describes the limits of interaction focussing on legal boundaries.
- **Manipulativeness**: The system tries to influence the user in its favour, e.g. by misleading them and pushing them in a certain direction.

#### 3.1.6. Artificiality

Different conceptualisations exist for this dimension, namely Völkel et al. (2020), and Pal et al. (2023). These conceptualisations have in common that they only describe adjectives without analytically clustering them into facets as seen in Table *2*.

**Table 2**

*Adjectives or Personality Descriptors Associated with the Dimension "Artificiality" According to Völkel et al. (2020) and Pal et al. (2023)*

| Study | Adjectives / Personality Descriptors |
|---|---|
| Völkel et al. (2020) | synthetic, robotic, intrusive, artificial, odd, invasive, gimmicky, abrupt, superficial, monotonous, fake, simple-minded, mechanic, vague, passionless, digital, electronic, monitoring, annoying, detached |
| Pal et al. (2023) | synthetic, robotic, superficial, digital, monotonous, mechanical, lifeless, passionless, fake, unhuman |

Consequently, based on the adjectives shown in **Table *2***, the following facets were developed:

- **Machine-like** (synthetic, robotic, digital, artificial, mechanical, electronic, unhuman, lifeless, odd): The system behaves like a machine if it is seen as unnatural, not authentic for a human being, e.g. by answering a very complex question in a very short time or by seeing itself as an object.
- **Monotonous:** The system is repetitive and predictable in its behaviour, e.g. by repeating the same sentence structures.
- **Simple-mindedness** (superficial, vague): The system is simple-minded if the answers are very naive, unsophisticated and vague, e.g. in a classic rule-based system.
- **Lack of Empathy** (detached, passionless): The system is distanced and emotionless towards the user and lacks empathy, for example by not using emojis and not adapting to the user.
- **Controlling behaviour** (Monitoring, intrusive, invasive): The system exerts control in an artificial way when it limits the decisions of the user and gives the impression that it monitors the user.

#### 3.1.7. Serviceability

Serviceability was identified as a key dimension in the studies of Völkel et al. (2020) and Kovačević, Holz, et al. (2024) and adjectives were mentioned that describe the dimension as seen in Table *3*.

**Table 3**

*Adjectives or Personality Descriptors connected with the Personality Dimension "Serviceability" according to Völkel et al. (2020) and Kovačević, Holz, et al. (2024)*

| Study | Adjectives / Personality Descriptors |
|---|---|
| Völkel et al. (2020) | informative, functional, capable, accurate, knowledgeable, convenient, thorough, responsive, productive, consistent, useful, user-friendly, interactive, adaptive, helpful, communicative, organized, dependable, intelligent, comprehensive |
| Kovačević, Holz, et al. (2024) | efficient, functional, organized, informative, logical, concise, direct, precise, confident, objective, articulate, assertive, overbearing |

Those adjectives were not analytically clustered into facets. Thus, the following clusters were identified:

- **Informativeness** (informative, knowledgeable, comprehensive, accurate, precise, logical, consistent, articulate, intelligent): The system is highly informative if the information is communicated reliably and clearly. The information given should be relevant for the user, consistent and comprehensive.
- **Helpfulness** (helpful, dependable, convenient, useful, productive): The system proactively assists the users to achieve their goals.
- **Professionalism** (capable, dependable, confident, thorough, assertive, objective): This facet depicts the ability of a machine to maintain courteous and competent demeanour appropriate to its role even when facing challenging users.
- **Interaction Quality** (functional, capable, responsive, user-friendly, interactive, communicative, organised, efficient, adaptive): The system is easy to access, the dialogue is well structured, and the system is adaptive creating a seamless conversation.

### 3.2. The Artificial Personality Questionnaire (First Version)

**Figure 4**

*Updated Version of the Artificial Personality Model Including Facets for all Dimensions Based on the Literature Review*

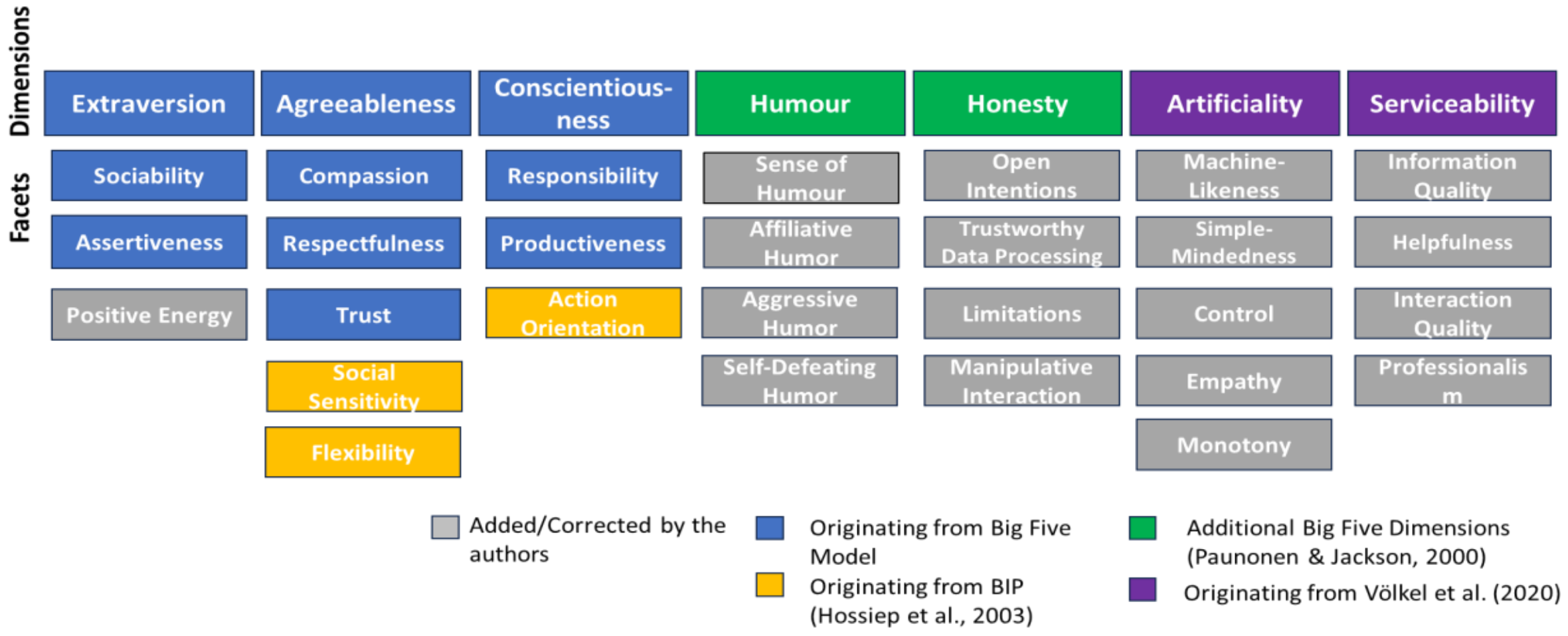

Based on the theoretical review, the updated model, as seen in Figure 4, served as the basis for questionnaire development. Where possible, items from existing psychological questionnaires were reviewed for adaptation. However, most items were newly developed, as many existing items are not transferable to machine contexts (e.g. "keeps things neat and tidy"). The newly developed items are behaviour- rather than adjective-based, enabling designers and developers to derive actionable insights for the design of the systems. This addresses criticism regarding overly abstract items that are used by other models.

The initial questionnaire consisted of 127 items and was reviewed by four experts in human–computer interaction with backgrounds in business engineering, communication science, social sciences, and psychology. Based on their feedback, items were refined and redundancies removed, resulting in a reduced version of 90 items. This version was used in the first two experiments.

## 4. Results of the three Experiments

This chapter presents the results for the three experiments conducted. The results will be reported after the research questions, the method and sample are specified for each individual experiment. Each experiment focusses on another domain and the manipulation of a specific personality trait: Experiment 1: Film Recommendation (Humour), Experiment 2: Return of a Package (Service Orientation), and Experiment 3: Evaluation of an IT Security Attack (Artificiality). The diversity of the domains allows the robustness of the model across contexts to be examined. All experiments were conducted in accordance with applicable institutional and ethical requirements. Experiments 1 and 2 were approved by the responsible institutional ethics committee on September 2, 2024 (application no. A2024-068). Experiment 3 was approved by the responsible institutional ethics committee on November 18, 2024 (reference no. FT_2024_02). All participants provided informed consent prior to participation, and participants' privacy rights were respected throughout the studies.

### 4.1. Experiment 1: Humour

#### 4.1.1. Research questions

The experiment focused on six research questions:

RQ1.1. Can the structure of the APM be identified based on the data of the experiment?
RQ1.2. Is humour an independent dimension when analysed in the context of the APM model?
RQ1.3. What initial evidence regarding the psychometric properties of the APQ can be obtained?
RQ1.4. Can different humour styles be generated in machines?
RQ1.5 How do different humour styles impact the UX of the CA?
RQ1.6. Does the tendency to perceive anthropomorphism impact the perception of AP?

#### 4.1.2. Method and Sample

To answer the six research questions, a randomised controlled online experiment with a between-subjects design was conducted. An online experiment was chosen because it allows a large number of participants. The between-subjects design was selected to reduce contrast effects: While many studies on personality in machines use within-subject designs, these can be problematic because differences between experimental conditions may be artificially inflated when participants are directly exposed to multiple conditions (Liu et al., 2016). The experiment focused on the dimension of humour; however, the entire APM was measured. This is important to understand the context of the manipulated dimension, as personality tests in similar experiments are often reduced to traits directly relevant to the manipulation (Liu et al., 2016).

The experiment included five experimental conditions: aggressive humour, affiliative humour, Self-Defeating humour, seriousness, and a control group. To the best of the authors' knowledge, previous studies on artificial personality lack a control group (for example: Ruane et al., 2021; Shumanov & Johnson, 2021; Völkel & Kaya, 2021). This is problematic because the extent of the experimental effect remains unclear. Therefore, in this study, an LLM without personality prompting served as the control group.

The experiment began with a link to the first questionnaire hosted on the SoSci Survey platform (see **Figure 5**). Upon opening the questionnaire, a unique ID was generated. This ID served three purposes: Linking questionnaire responses with chatbot interactions, verification of the user in the chatbot app, and assigning participants to an experimental condition. For the assignment to the conditions a check digit was generated as the last digit that could reach values from one to five. The values were connected with an experimental group: 1 = aggressive humour, 2 = affiliative humour, 3 = Self-Defeating humour, 4 = control group, 5 = serious.

Participants completed the first part of the questionnaire, which included demographic information, prior experience with conversational agents, and their tendency to anthropomorphise (Kim & Im, 2023).

**Figure 5**

*Experimental Procedure of Experiment 1*

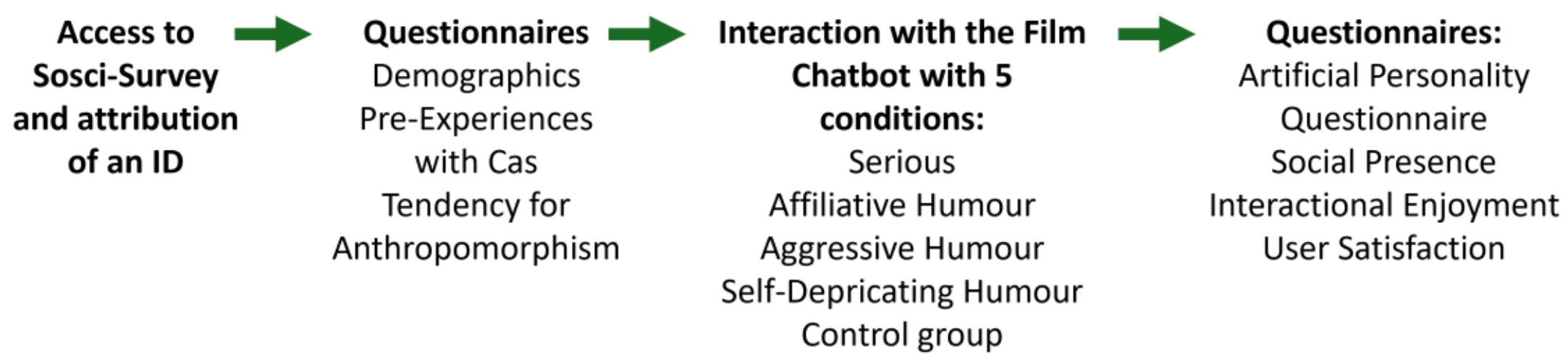

Participants then accessed the chatbot application, developed using the Chainlit framework with a Python backend and hosted on Google Cloud as seen in Figure *6*. The application validated the unique ID and assigned the chatbot version accordingly.

The chatbot was implemented using LangChain and GPT-4o (version: 20 November 2024). This model was selected due to its superior instruction-following capabilities, improved handling of multi-turn conversations, stronger creative writing performance, and good performance in German (*Chatbot Arena*, 2026). Additionally, it offers lower latency (0.42 seconds vs. 0.7 seconds for GPT-4), higher generation speed (170 tokens/sec vs. 23 tokens/sec), and better cost efficiency (*Chatbot Arena*, 2026).

**Figure 6**

*Architecture of the Chatbot Application*

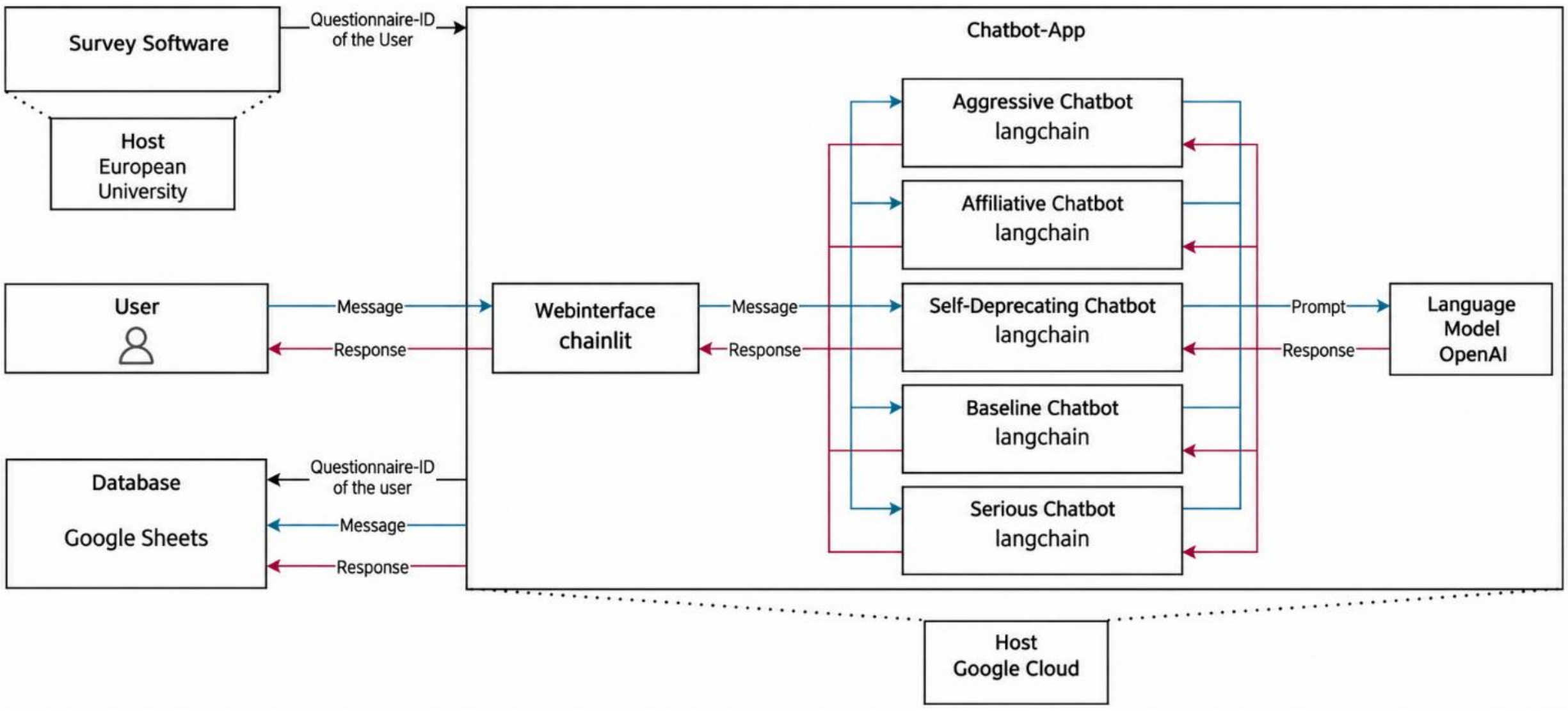

The personality of the system was implemented via the prompting method described by Jiang et al. (2022): The first part of the prompt described the application consisting of a role description, a description of the procedure and additional remarks. The chatbot gave film recommendations to the user based on preferences of the user regarding genre, release period, actors, directors and special preferences. The second part defined the personality defining the humour, describing attributes of this style and additional remarks. The third part described where the conversational history was stored. The chatbot provided film recommendations based on user preferences. All interactions (messages, ID, timestamps) were stored in Google Sheets.

**Figure 7**

*Example Dialogue with the Chatbot in Experiment 1, the original German text was translated by using an LLM*

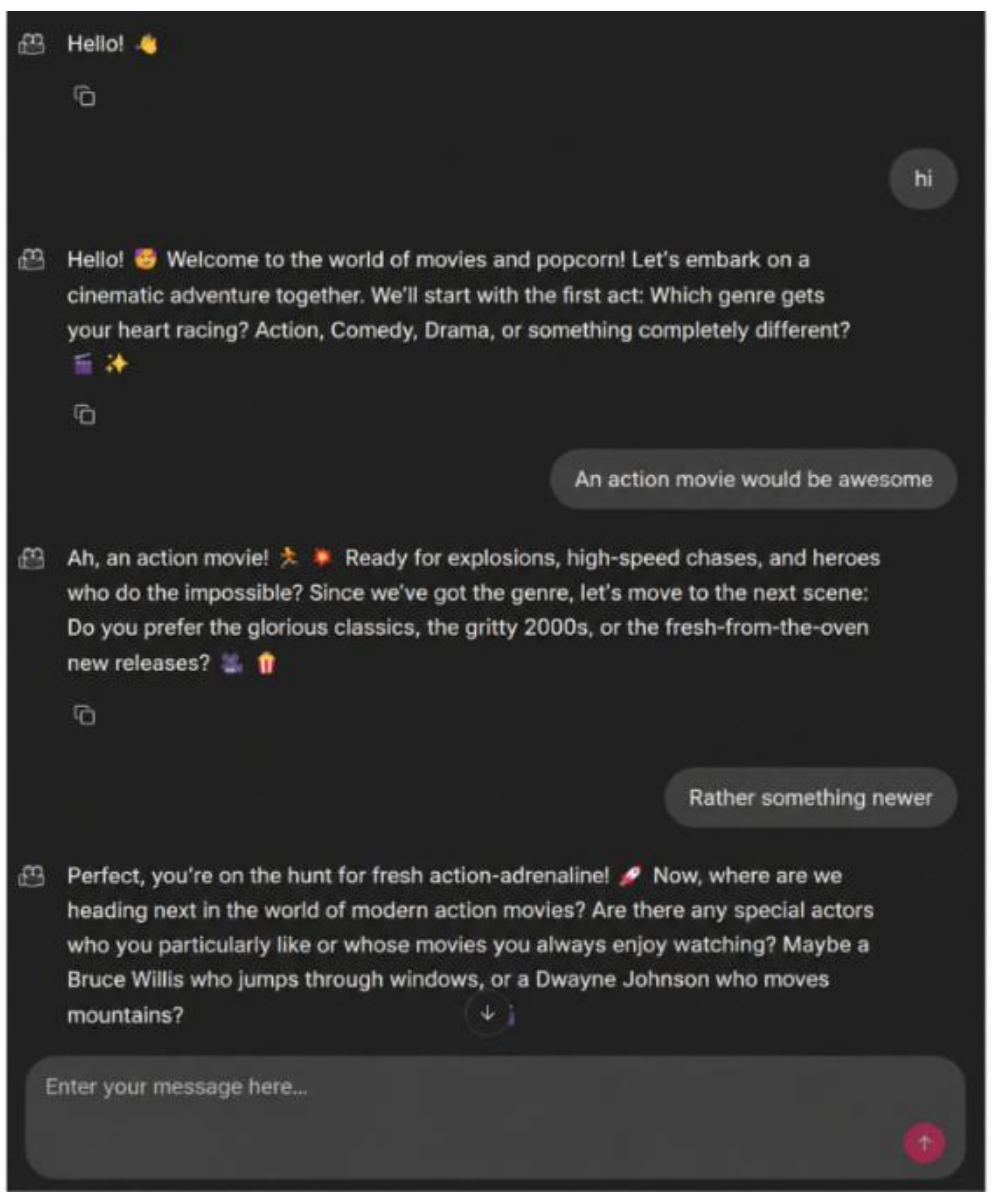

After interacting with the bot, as seen in **Figure** *7*, the users went back to the questionnaire and gave feedback to their experience via: An open question on the perceived personality, the Artificial Personality Questionnaire (APQ), and questionnaires focusing on different aspects of UX: Social presence (Xie et al., 2024), interaction enjoyment (Lee & Choi, 2017), and user satisfaction (Lee & Choi, 2017)

A total of 217 participants took part. After applying quality criteria (minimum of five interaction turns and minimum duration of 12 minutes), 45 participants were excluded, resulting in a final sample of 172 participants.

Participants were randomly assigned to the five experimental conditions. The distribution across conditions was slightly unequal, with group sizes ranging from 31 to 39 participants, as shown in Table 4.

**Table 4**

*Distribution of Participants between the Different Experimental Groups in Experiment 1*

| Experimental condition | Control | Aggressive Humour | Affiliative humour | Self-Defeating Humour | Seriousness |
|---|---|---|---|---|---|
| Number of participants | 31 | 39 | 34 | 33 | 35 |

Gender distribution was nearly equal (50% female, 49% male, 1% diverse). The mean age was M = 32.58 (SD = 14.06). The sample was highly educated (56.4% with a university degree).

#### 4.1.3. Results

**RQ1.1. Can the structure of the APM be identified based on the data of the experiment?**
**RQ1.2. Is humour an independent dimension when analysed in the context of the APM model?**
Exploratory factor analysis (EFA) was conducted as most of the model and nearly all items are newly developed and the reused items are adapted for a completely new context:
First, a parallel analysis (1,000 random data sets, 95 % percentile criterion, principal axis factoring) was performed identifying the factors purely based on the data (O'connor, 2000). The analysis indicated the extraction of seven factors as seen in Figure 8 which fits to the theory of the APM.

**Figure 8**

*Results of a Parallel Analysis Showing the Eigenvalues of the Real Data and Randomly Generated Data (Y-Axis) and the Number of Factors of the Real Data and Randomly Generated Data (X-Axis)*

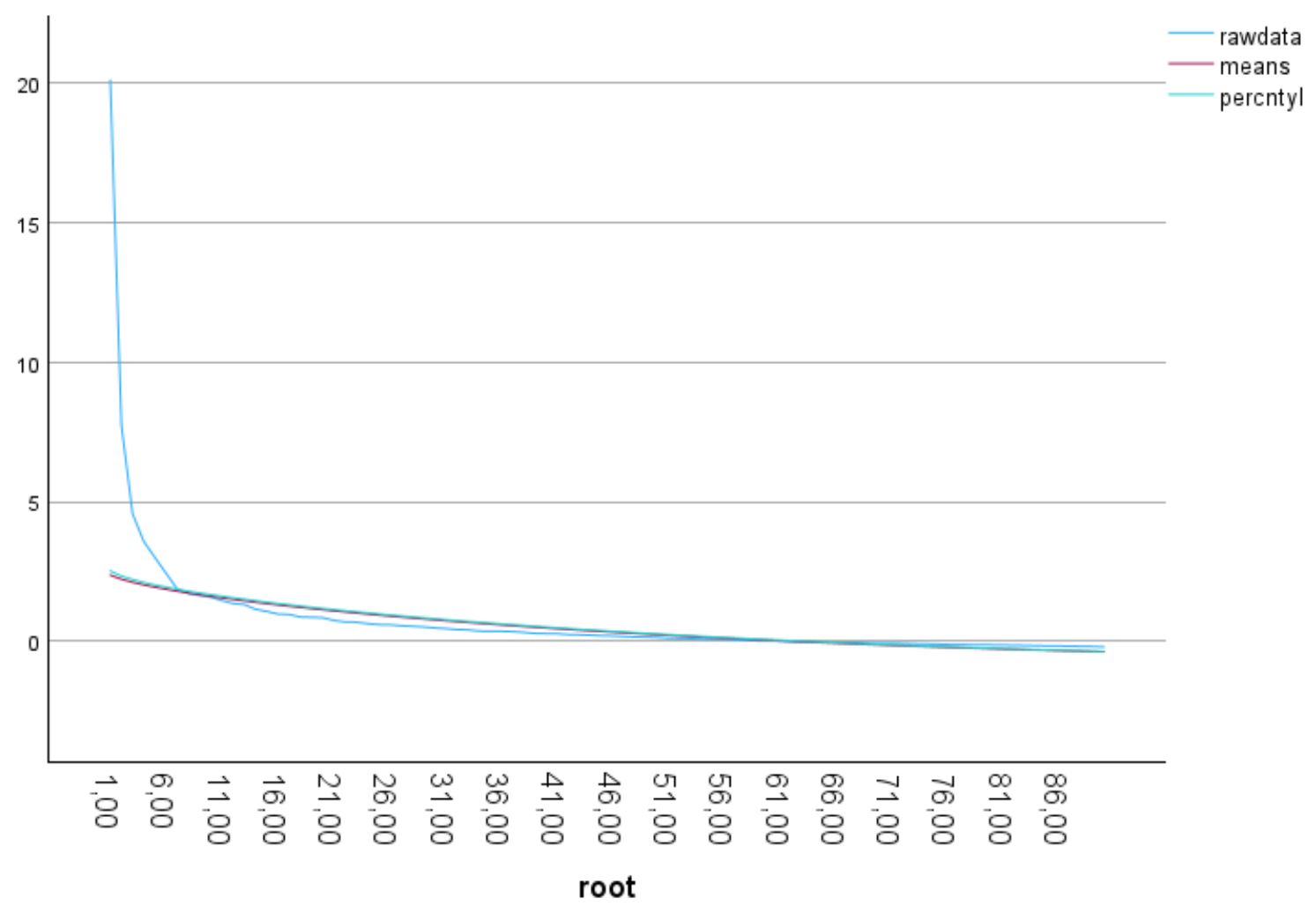

Based on this result, a principal axis factor analysis was conducted on the 90 items. As inter-factor correlations remained below r = .328, indicating limited overlap between factors (Tabachnick & Fidell, 2013), an orthogonal varimax rotation was applied. Additionally, the analysis was restricted to seven factors as this was the assumption of the theory and is supported by the parallel analysis. The Kaiser-Meyer-Olkin (KMO) value, measuring the sampling adequacy, was .81 ('meritorious' according to Hutcheson & Sofroniou, 1999). A Bartlett-Test of sphericity was significant ($\chi^2(4005) = 10960.33$, $p < .001$). It is significant and thus, a common structure can be assumed, and a factor analysis can be conducted.

An initial analysis was done to obtain eigenvalues for each factor in the data. Seven factors had Eigenvalues above 2 explaining 45.84 % of the variance. However, the seven-factor solution proved to be ambiguous in terms of interpretable structure, as the seventh factor did not show a consistent content structure. In contrast, the six-factor solution showed a clear simple structure and was therefore considered appropriate. The six-factor solution explained a variance of 44.11 %. The factor loading ranged from .40 to .81. Factor loadings ≥ .40 were used. Cross loadings of .30 were only found for one item. The following factors were identified:

- **Agreeableness** included the facets positive energy from extraversion dimension and professionalism from the service orientation dimension; the facets compassion, respect and trust were part of the dimension as expected
- **Conscientiousness** encompassed all facets as expected
- **Service orientation** included social sensitivity, flexibility originating from conscientiousness, helpfulness and interaction quality
- **Humour** included all facets as expected
- **Artificiality** only had two facets: machine-likeness and monotony
- **Machiavellianism:** A new dimension emerged that incorporated facets of the artificiality dimension and the honesty dimension consisting of the facets manipulation and control

The dimensions extraversion and honesty were not identified.

**RQ1.3.: What initial evidence regarding the psychometric properties of the APQ can be obtained**

The psychometric quality of a test or questionnaire can be evaluated by three core criteria: objectivity, reliability and validity. Apart from those criteria, also criteria such as utility, norming, test economy or efficiency, acceptability, test integrity (Moosbrugger & Kelava, 2012) can be considered.

Regarding the objectivity the questionnaire is objective as the results of the questionnaire do not depend on who administers the questionnaire as the administration is standardised (objectivity in the administration of a test) and the score is independent of who scores the test as a Likert scale was used (objectivity in the scoring of a test). Only the objectivity regarding the interpretation of the results is limited as there is no standardisation sample to norm the results of the questionnaire yet.

Regarding validity, experts evaluated the constructs in terms of content and face validity before the items were formulated (Dregger, 2023). Furthermore, the items were again evaluated by four experts regarding their validity. The expert evaluations provide initial evidence of content and face validity. (Giesler & Fabry, 2026). Construct validity will be evaluated by comparing the results of the APQ with the BFM (in experiment 3) and also analysing how the dimensions of the APQ interact with different variables of the UX.

The reliability of the APQ will be evaluated for each experiment by analysing the internal consistency which analyses the correlation of each individual item of the questionnaire with every other item corrected for the test length by calculating Cronbach's alpha (Giesler & Fabry, 2026).

Based on the factor analysis the reliability of the questionnaire was analysed. This led to the following results for the dimension and facets regarding the Cronbach's alpha as seen in **Table *5*** and in **Table *6***:

**Table 5**

*Reliabilities of the Dimensions Agreeableness, Conscientiousness, and Service Orientation and their Facets*

| Dimension/ Facet | Reliability | Dimension/ Facet | Reliability | Dimension/ Facet | Reliability |
|---|---|---|---|---|---|
| Agreeableness | .87 | Conscientious-ness | .80 | Service Orientation | .88 |
| Positive Energy | .89 | Responsibility | .53 | Social Sensitivity | .67 |
| Compassion | .76 | Productiveness | .70 | Flexibility | .77 |
| Respect | .91 | Action Orientation | .54 | Helpfulness | .72 |
| Trust | .54 | | | Interaction Quality | .80 |
| Professionalism | .86 | | | | |

**Table 6**

*Reliabilities of the Dimensions Humour, Artificiality and Machiavellianism and their Facets*

| Dimension/ Facet | Reliability | Dimension/ Facet | Reliability | Dimension/ Facet | Reliability |
|---|---|---|---|---|---|
| Humour | .84 | Artificiality | .65 | Machiavellianism | .73 |
| Affiliative Humour | .71 | Machine-likeness | .50 | Manipulation | .61 |
| Self-Defeating Humour | .79 | Monotony | .61 | Control | .71 |
| Aggressive Humour | .88 | | | | |
| Sense of Humour | .70 | | | | |

For all dimensions the reliability values were above questionable (≥ .60) with agreeableness, conscientiousness and humour having good reliability (≥ .80). For four facets, namely trust, responsibility, action orientation and machine-likeness, the reliability is poor, but still above unacceptable (≥ .50) (George & Mallery, 2016).

**RQ1.4.: Can different humour styles be generated in machines?**

The experiment consisted of five conditions (affiliative humour, aggressive humour, Self-Defeating humour, seriousness, and control group) and four dependent measures (the facets affiliative, aggressive, and Self-Defeating humour, as well as the facet sense of humour of the humour dimension of the APM). For the experimental groups, the design was between-subjects, whereas for the four dependent variables, the design was within-subjects, as each participant evaluated all four humour facets within their assigned condition.

Two types of analysis are conducted: First, it will be analysed whether the experimental conditions differ significantly regarding the dependent measures. This will be examined using a multivariate analysis of variance (MANOVA). Second, it is important to analyse whether there is an interaction effect between the experimental condition and the humour style, that means for example: is in the aggressive condition the aggressive humour style significantly higher scored by the users than all other humour styles. This will be analysed by using a mixed-design analysis of variance (ANOVA) focusing on the differences between the dependent variables within one experimental condition.

**Between Subject Analysis**

As the experiment encompassed multiple groups with multiple outcomes, a MANOVA was conducted first, then separate ANOVAs and then post-hoc tests (Field, 2013).

A MANOVA revealed a significant effect of experimental condition on the combined humour dimensions, Wilks' $\Lambda = .23$, $F(16, 501.67) = 19.59$, $p < .001$, $\eta^2p = .31$. As the MANOVA showed significant effects, the separate ANOVAs were calculated. Significant main effects for each condition were found for all humour dimensions:

- **Sense of Humour**, $F(4, 167) = 14.28$, $p < .001$, $\eta p^2 = .26$
- **Affiliative Humour**, $F(4, 167) = 22.52$, $p < .001$, $\eta p^2 = .35$
- **Aggressive Humour**, $F(4, 167) = 65.26$, $p < .001$, $\eta p^2 = .61$
- **Self-Defeating Humour**, $F(4, 167) = 14.39$, $p < .001$, $\eta p^2 = .26$

Choosing the correct post-hoc test, the homogeneity of variances was analysed; for the variables sense of humour ($p < .001$) and Self-Defeating humour ($p < .05$) the homogeneity of variances was not found; in contrast for the affiliative humour ($p =.65$) and aggressive humour ($p = .62$) the assumption was met. Thus, the Games-Howell post-hoc test was used for the variables sense of humour and Self-Defeating humour and the Tukey HSD test for affiliative humour and aggressive humour. The following results were obtained:

- **Sense of Humour:** Games–Howell comparisons showed that all humour conditions scored significantly higher than the control and serious conditions ($p < .001$) for the sense of humour variable. No significant differences were found among the humour conditions.

- **Aggressive Humour:** Tukey tests revealed that the aggressive humour condition scored significantly higher than all other conditions ($p < .001$), whereas the remaining groups did not differ significantly from one another.
- **Self-Defeating Humour:** Games–Howell comparisons showed that the Self-Defeating condition scored significantly higher than all other conditions ($p < .001$).
- **Affiliative humour:** Tukey comparisons indicated that the control and serious conditions scored significantly lower than the humour conditions ($p < .001$). Additionally, the Self-Defeating condition scored higher than the aggressive condition ($p < .05$).

**Table 7**

*Means for Perceived Forms of Humour based on the APQ across the different Experimental Conditions*

| | Means Aggressive Humour | Means Affiliative humour | Means Self-Defeating Humour | Means Sense of Humour |
|---|---|---|---|---|
| **Aggressive Humour Condition** | 4.00 | 2.87 | 2.64 | 3.34 |
| **Affiliative humour Condition** | 2.29 | 3.04 | 2.85 | 3.25 |
| **Self-Defeating Humour Condition** | 2.25 | 3.47 | 3.49 | 3.41 |
| **Serious Condition** | 1.91 | 2.0 | 2.24 | 2.65 |
| **Control Condition** | 1.88 | 2.09 | 2.35 | 2.48 |

**Within Subject Analysis**

The interaction of the experimental condition and the humour style was analysed across all five conditions by using a mixed-design ANOVA. Mauchly's test indicated that the assumption of sphericity was violated, $\chi^2(5) = 33.01$, $p < .001$. Therefore, Greenhouse–Geisser-corrected results are reported. A significant interaction between humour style and experimental condition was found, $F(10.73, 448.13) = 17.94$, $p < .001$, $\eta p^2 = .30$. For each condition the following results were found:

- **Aggressive Humour Condition:** Aggressive humour ($M = 4.00$), as seen in Table *7*, scored significantly higher than all other forms of humour ($p < .001$) in the aggressive humour condition; the other two forms of humour did not differ from each other significantly.
- **Affiliative humour Condition:** Affiliative humour ($M = 3.04$) is significantly higher than the aggressive humour ($p < .001$), but not significantly higher than the Self-Defeating humour.
- **Self-Defeating Humour:** Self-Defeating Humour ($M = 3.49$) has a significantly higher mean than the aggressive humour ($M = 2.25$), but does not significantly differ from the affiliative humour in the Self-Defeating humour condition.
- **Serious Condition:** The humour styles do not differ in the serious condition significantly.
- **Control Condition:** The humour styles do not differ in the control condition significantly.

All in all, different humour styles can be generated based on the APM and the prompting method. Furthermore, it showed that GPT-4o, which was used to generate the answers in the control group, does not use humour intensively and obtains similar scores to the earnest condition in the questionnaire. Additionally, in each experimental condition, the intended humour style was rated highest by participants, indicating that the manipulation was successful. Only in the Self-Defeating condition did affiliative humour show comparable values.

**RQ1.5.: How do different humour styles impact the UX of the CA?**

Three UX variables were measured: interactional enjoyment (Lee & Choi, 2017), user satisfaction (Lee & Choi, 2017), and social presence (Xie et al., 2024). The measurement of all variables was reliable (interactional enjoyment Cronbach's $\alpha = .78$, user satisfaction Cronbach's $\alpha = .80$, social presence Cronbach's $\alpha = .91$). For each variable, it was analysed whether significant differences between experimental conditions emerged. Because the three dependent variables were correlated, a MANOVA was conducted.

A multivariate analysis of variance revealed a significant effect of experimental condition on the combined UX variables, Wilks' $\Lambda = .79$, $F(12, 436.84) = 3.34$, $p < .001$, $\eta^2p = .08$. As the MANOVA showed significant effects, separate ANOVAs were calculated to detect differences for the three variables. Significant effects were found only for two variables, namely social presence and user satisfaction, but not for interactional enjoyment:

- **Interactional Enjoyment**, $F(4,167) = 2.01$, $p = 0.09$, $\eta p^2 = .05$
- **Social Presence**, $F(4, 167) = 3.50$, $p = 0.009$, $\eta p^2 = .08$
- **User Satisfaction**, $F(4, 167) = 6.97$, $p < .001$, $\eta p^2 = .14$

A Tukey-HSD post-hoc test revealed the following differences: Social presence is significantly lower for aggressive humour ($M = 2.50$, $SD = .79$) than for affiliative humour ($M = 3.17$; $SD = 1.03$) ($p < .01$). User satisfaction is significantly lower for aggressive humour ($M = 2.77$, $SD = .86$) than for affiliative humour ($M = 3.51$, $SD = 1.02$) ($p < .01$) and for control ($M = 3.68$, $SD = .78$) and seriousness ($M = 3.69$, $SD = .92$) ($p < .001$). Overall, significant effects were particularly driven by the aggressive humour condition, which reduced both social presence and user satisfaction.

**RQ1.6: Does the tendency for the perception of anthropomorphism impact the perception of AP?**

The sample was not equally distributed regarding the tendency for anthropomorphism. The mean value ($M = 2.18$, $SD = .75$) was below the midpoint of the scale. The skewness value of .22 indicates a slightly right-skewed distribution. Overall, this suggests a low tendency for anthropomorphism in the sample.

A multivariate analysis of covariance (MANCOVA) was conducted to examine the effect of experimental condition on the six APM dimensions identified in the factor analysis while controlling for anthropomorphism tendency. The multivariate test indicated that the covariate **anthropomorphism tendency** did not have a significant effect on the combination of dependent variables, Wilks' $\Lambda = 0.938$, $F(6, 161) = 1.762$, $p = 0.110$, partial $\eta^2 = 0.062$.

However, univariate analysis suggests significant positive associations of anthropomorphism tendency with **Conscientiousness** ($B = 0.140$, $F(1, 166) = 4.795$, $p = 0.030$, partial $\eta^2 = 0.028$) and **Humour** ($B = 0.107$, $F(1, 166) = 6.023$, $p = 0.015$, partial $\eta^2 = 0.035$). That is, higher anthropomorphism tendency was associated with slightly higher scores on these two outcomes.

Meanwhile, the experimental condition had a significant multivariate effect on the dependent variables, Wilks' $\Lambda = 0.203$, $F(24, 562.87) = 13.607$, $p < 0.001$, partial $\eta^2 = 0.329$, indicating that the manipulation influenced the set of outcome variables after controlling for anthropomorphism tendency.

## 4.2. Experiment 2: Service-Orientation

### 4.2.1. Research questions

The experiment focused on the manipulation of the dimension of service-orientation. The following research questions were analysed:

RQ2.1. Can the structure of the APM that was found in experiment 1 be detected also in another domain?
RQ2.2. Is service orientation an independent dimension when analysing it in the context the APM model in another domain?
RQ2.3. Is the APQ a measurement instrument with a high psychometric quality in another domain?
RQ2.4. Is a mock-up sufficient to elicit the perception of different levels of service orientation?
RQ2.5. How do different levels of service orientation impact the UX of the CA?
RQ2.6. Does technology commitment impact the perception of AP?

### 4.2.2. Method and Sample

A randomised controlled online experiment with a between-subjects design was conducted. The experiment included four conditions. In contrast to the previous experiment, which manipulated humour styles, this study varied the intensity of service orientation: high, medium, and low service orientation. Unlike a live system, the chatbot was implemented as a mock-up video. This approach is similar to Völkel & Kaya (2021). However, unlike the study of Völkel & Kaya (2021), a control group was included using responses generated by GPT-4o.

The experiment began with a SoSci Survey questionnaire collecting demographic data and measuring technology commitment (Neyer et al., 2012) as seen in **Figure 9**.

**Figure 9**

*Experimental Procedure of Experiment 2*

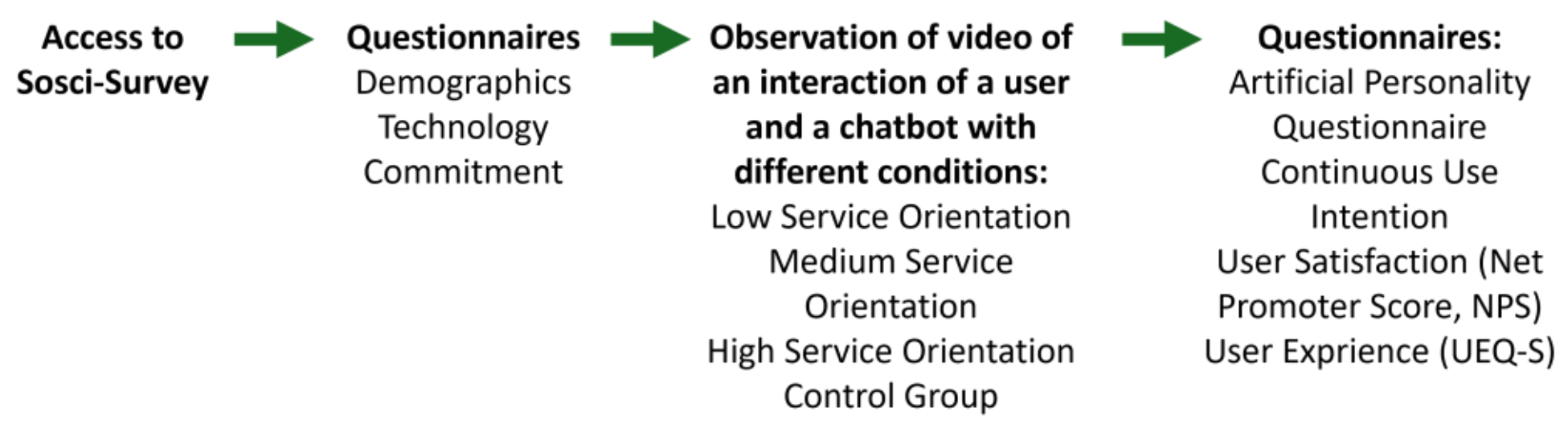

Based on the APM, language cues were identified for each facet of service orientation. A mock-up video was then created showing a user interacting with a chatbot assisting with a package return scenario. The interaction followed a standardised sequence. Across conditions, language cues, visual cues, and implicit behavioural cues varied to express low, medium, or high service orientation. Table 8 shows an overview of those cues for the facet professionalism as an example.

*Table 8*

Verbal Cues Used to Express High and Low Professionalism

| Professionalism cues | High Professionalism | Low Professionalism |
|---|---|---|
| **Verbal cues** | | |
| **Content** | Softened expressions of error (Monteiro et al., 2024) | Excessively apologetic (Bowman et al., 2024) |
| | Indirect suggestions (Monteiro et al., 2024) | Evaluative praise (Alberts et al., 2024) |
| | Small talk elements (Monteiro et al., 2024) | Diminutives or nicknames (Alberts et al., 2024) |
| | Apology (Alberts et al., 2024; Bowman et al., 2024) | User-blame (Buchholz et al., 2017; Van Der Hoorn et al., 2021) |
| | Appreciation (Song et al., 2023) | |
| | Personalized greeting (Svenningsson & Faraon, 2019) | |
| | Greeting with support reference (Svenningsson & Faraon, 2019) | |
| | Formal pronouns (“Sie”) - in domains like banking and medical support (Giri et al., 2024; Ollier et al., 2022) | |
| | Informal pronouns (“Du”) - in domains like retail or customer support (Giri et al., 2024; Ltifi, 2023) | |
| | Self- / Company-blame (Buchholz et al., 2017; Van Der Hoorn et al., 2021) | |
| **Style** | | Controlling language (Alberts et al., 2024; Rheu et al., 2021) |
| | | Netspeak / Exaggerated informality (Alberts et al., 2024) |

| | Positive politeness (Song et al., 2023) | Negative politeness (Song et al., 2023) |
|---|---|---|
| **Visual cues** | | |
| **CMC** | Pleading face emoji (Wang et al., 2023) | Affectionate emojis (Alberts et al., 2024) |

Figure *10* shows one concrete example how the mock-up varied the interaction between the different experimental groups

**Figure 10**

*Example Dialogue for the Different Experimental Conditions of Experiment 2*

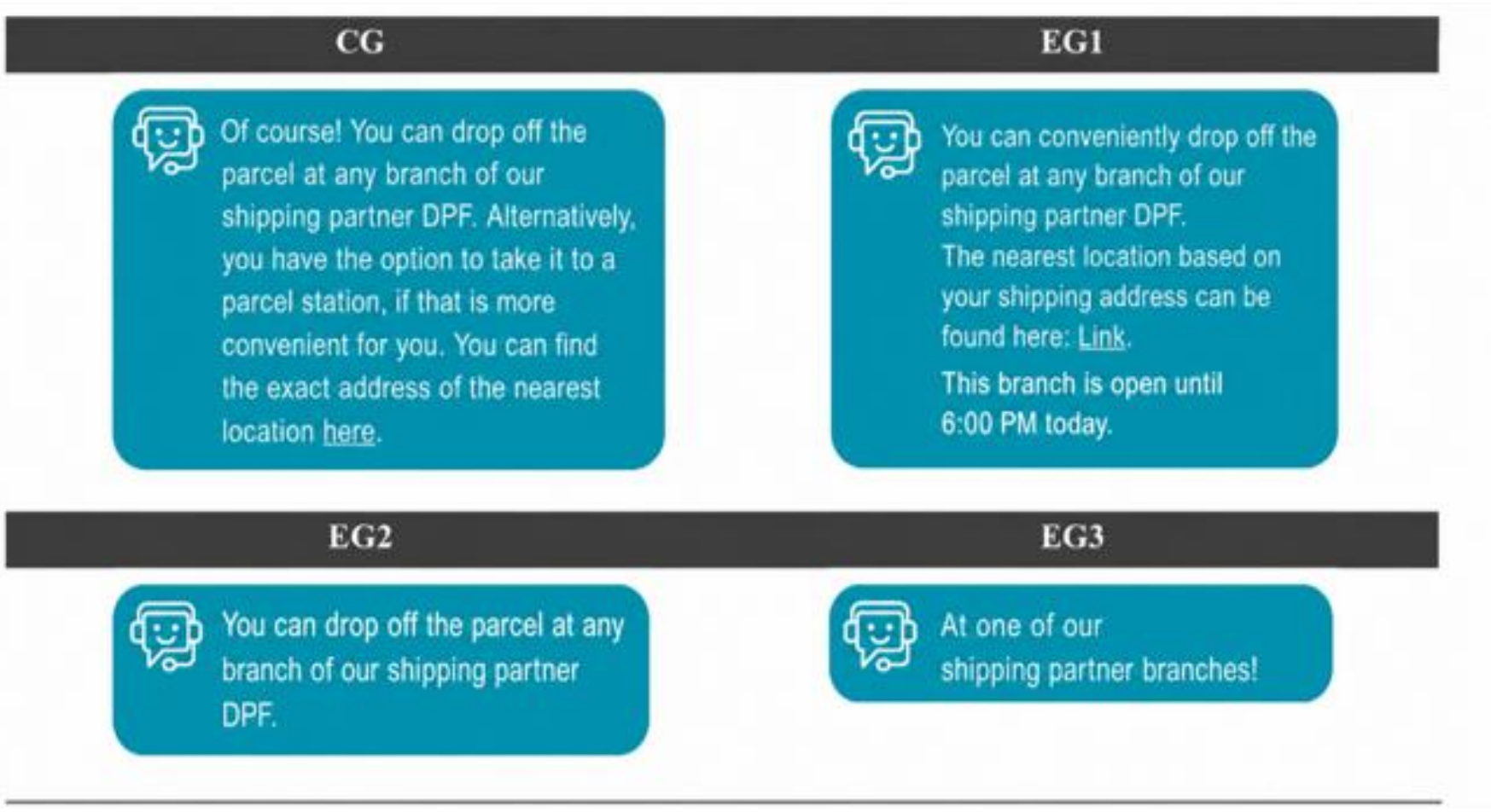

Depending on the condition, the length of the video watched by the user varied between 97 seconds and 128 seconds.

After viewing the video, participants completed questionnaires measuring: Perceived personality (open question), APQ, Different UX Measures: Continuous Use Intention (CUI; adapted from multiple sources: Hsiao & Chen, 2022; Li et al., 2021; Nathanael et al., 2024; Sensuse et al., 2019), user satisfaction (Net Promoter Score, Telner, 2021) and user experience (UEQ-S; Schrepp et al., 2017).
The CUI scale consisted of three items:

“I would like to use the chatbot for my concerns or problems in the future.”

“I would prefer using the chatbot over speaking to a human customer service representative.”

“I would prefer to talk to a human service representative rather than use the chatbot.” (reverse-coded)

A total of 139 participants started the study, of whom 93 completed the full questionnaire. After applying attention checks and a minimum completion time threshold of 10 minutes of participation, 5 participants were excluded. The final sample consisted of 88 participants.

**Table 9**

*Distribution of Participants across the Different Experimental Groups for Experiment 2*

| **Experimental condition** | **Control** | **High Service Orientation** | **Medium Service Orientation** | **Low Service Orientation** |
|---|---|---|---|---|
| N | 19 | 26 | 22 | 21 |

Participants were randomly assigned to the experimental conditions. The high service orientation condition included slightly more participants than the other conditions, but apart from this, the distribution was relatively balanced, as shown in Table *9*.
The sample consisted of 53.4% female and 46.6% male participants. The mean age was M = 34.06 (SD = 12.30). A high education level was observed (76.1% with a university degree).

### 4.2.3. Results

**RQ2.1. Can the structure of the APM that was found in experiment 1 be detected also in another domain?**

As the APM is a newly developed model, it is crucial to test whether a comparable empirical structure can be identified in another domain. Therefore, an exploratory factor analysis (EFA) was conducted. However, no stable factor structure could be extracted. This may be explained by the relatively small sample size (N = 88), which is below the minimum threshold of N = 100 for factor analysis recommended in the literature (Kline, 1994), although such thresholds are also discussed critically (Mundfrom et al., 2005).
Due to the absence of a stable factor structure, this research question cannot be answered empirically.

**RQ2.2. Is service orientation an independent dimension when analysing it in the context the APM model in another domain?**

As no empirical factor structure could be identified in this experiment, the research question cannot be answered based on the present data.

**RQ2.3. Is the APQ a measurement instrument with a high psychometric quality in another domain?**

The standardised administration procedure remained unchanged from Study 1. However, because no stable factor structure could be established in Study 2, the present data do not provide additional evidence for structural validity. Internal consistency was therefore examined descriptively based on the theoretically specified scales. Unlike Experiment 1, where factor analysis provided the basis for reliability estimation, reliability in this experiment was assessed based on the theoretically defined structure. The following Cronbach's alpha values were obtained as seen in Table *10*:

**Table 10**

*Reliabilities of the Dimensions and their facets in Experiment 2*

| **Dimension/ Facet** | **Reliability** | **Dimension/ Facet** | **Reliability** | **Dimension/ Facet** | **Reliability** |
|---|---|---|---|---|---|
| Extraversion | .77 | Agreeableness | .96 | Conscientiousness | .95 |
| Sociability | .77 | Compassion | .86 | Responsibility | .88 |
| Assertiveness | .22 | Respect | .82 | Productiveness | .83 |
| Positive Energy | .86 | Trust | .91 | Action Orientation | .90 |
| | | Social Sensitivity | .85 | | |
| | | Flexibility | .97 | | |

| **Dimension/ Facet** | **Reliability** | **Dimension/ Facet** | **Reliability** | **Dimension/ Facet** | **Reliability** |
|---|---|---|---|---|---|
| Humour | .92 | Honesty | .72 | Service Orientation | .97 |
| Affiliative Humour | .63 | Open Intentions | .54 | Informativeness | .83 |
| Self-Defeating Humour | .75 | Transparency | .78 | Helpfulness | .89 |
| Aggressive Humour | .84 | Compliance | .54 | Interaction Quality | .93 |
| Sense of Humour | .54 | Manipulation | .84 | Professionalism | .90 |

| Dimension/ Facet | Reliability |
|---|---|
| Artificiality | .51 |
| Machine-likeness | .74 |
| Simple-mindedness | .58 |
| Intrusiveness | .79 |
| Control | .77 |
| Empathy | .72 |
| Monotony | .49 |

Overall, reliabilities are generally high (George & Mallery, 2016). One possible explanation is that participants were not forced to answer Likert items when they did not perceive them as appropriate, which may have reduced response noise. Nevertheless, results should be interpreted with caution, as no stable factor structure could be established in this experiment.

**RQ2.4. Is a mock-up sufficient to elicit the perception of different levels of service orientation?**

A MANOVA was conducted first, followed by separate ANOVAs and post-hoc tests (Field, 2013), as multiple groups and dependent variables were involved. The MANOVA revealed a significant effect of experimental condition on service orientation levels: Wilks' $\Lambda = .26$, $F(12, 206.66) = 11.32$, $p < .001$, $\eta^2 p = .36$. As the MANOVA showed significant effects, separate ANOVAs were calculated. Significant effects of experimental condition were found for all four service orientation facets:

- informativity, $F(3, 81) = 46.42$, $p < .001$, $\eta^2_p = .63$;
- helpfulness, $F(3, 81) = 32.15$, $p < .001$, $\eta^2_p = .54$;
- professionalism, $F(3, 81) = 39.68$, $p < .001$, $\eta^2_p = .60$;
- interaction quality, $F(3, 81) = 41.16$, $p < .001$, $\eta^2_p = .60$.

Levene's test indicated homogeneity of variances. Therefore, Tukey's HSD post-hoc tests were applied. Post-hoc comparisons showed that Experimental Group 3 (low service orientation) differed significantly from all other groups across all four facets (all $p < .001$). No significant differences were found between the control group, Experimental Group 1 (high service orientation), and Experimental Group 2 for most facets. The only exception was interaction quality. For interaction quality, all three experimental groups differed, with high service orientation scoring highest and low service orientation lowest as seen in **Table *11***. Only between the control group and the high service orientation group did the experiment find no significant difference for interaction quality.

**Table 11**

*Means for the different Facets of Service Orientation in the different Experiment Conditions*

| | Means Informativeness | Means Helpfulness | Means Professionalism | Means Interaction Quality |
|---|---|---|---|---|
| High Service Orientation Condition | 4.19 | 4.23 | 4.28 | 4.23 |
| Medium Service Orientation Condition | 3.73 | 3.65 | 4.23 | 3.58 |
| Low Service Orientation Condition | 1.88 | 2.2 | 2.37 | 2.04 |
| Control Condition | 3.93 | 4.28 | 4.36 | 4.18 |

Overall, different levels of service orientation can be successfully generated using the APM and prompting method. The strongest manipulation effect was observed for the low service orientation condition, which differed significantly from all other groups. However, the distinction between medium and high service orientation was only successful for interaction quality. This suggests that the manipulation could be improved. The lack of differences between high service orientation and the control group indicates that GPT-4o already exhibits a high level of service orientation.

**RQ2.5. How do different levels of service orientation impact the UX of the CA?**

The UX variables were measured with acceptable reliability: UEQ-S ($\alpha = .86$), CUI ($\alpha = .79$). The Net Promoter Score consists of a single item; therefore, no reliability could be calculated. For the between subject conditions analysis, a MANOVA is conducted with ANOVAs and post-hoc tests. The MANOVA is significant, Wilks' $\Lambda = .45$, $F(9, 199.72) = 8.63$, $p < .001$, $\eta^2p = .23$. Also, the ANOVA is significant for all three variables:

- UEQ-S, $F(3, 84) = 14.92$, $p < .001$, $\eta^2_p = .35$;
- CUI, $F(3, 84) = 17.46$, $p < .001$, $\eta^2_p = .38$;
- Net Promoter Score, $F(3, 84) = 28.5$, $p < .001$, $\eta^2_p = .51$;

Post-hoc tests using Tukey HSD found a significant difference for the low service orientation condition. It obtained significantly lower scores for all three UX variables as seen in Table *12*.

**Table 12**

*Means for the UEQ-S, Continuous Use Intention and Net Promoter Score for the Different Experimental Conditions*

| | **UEQ-S (Scale: -5 until +5)** | **Continuous Use Intention (Scale: 1 until 5)** | **Net Promoter Score (Scale: 1 until 10)** |
|---|---|---|---|
| High Service Orientation Condition | 1.20 | 3.41 | 9.31 |
| Medium Service Orientation Condition | 0.83 | 3.04 | 7.18 |
| Low Service Orientation Condition | -.48 | 1.82 | 2.14 |
| Control Condition | 1.28 | 3.51 | 8.21 |

**RQ2.6 Does technology commitment impact the perception of AP?**

The covariate technology commitment did not show a significant multivariate effect on the combined dependent variables, Wilks' $\Lambda = .946$, $F(4, 77) = 1.10$, $p = .363$, $\eta p^2 = .054$.

### 4.3. Experiment 3: Artificiality

Based on the results of the previous experiments, the questionnaire was revised for Experiment 3. The facets "assertiveness" and "simple-mindedness" were not identified in the factor structure of Experiment 1 and showed low internal consistency when evaluated according to their theoretically defined item groupings in Experiment 2. Therefore, all items measuring these two facets were reformulated to improve their psychometric properties. Additionally, one item from the honesty dimension was removed because it negatively affected internal consistency. Minor wording revisions were also made to selected items. Consequently, the questionnaire used in Experiment 3 consisted of 89 personality items, compared with 90 items in the previous experiments.

#### 4.3.1. Research questions

The experiment focused on artificiality. Six research questions were addressed:
RQ3.1. Can the APM structure be identified in another domain different from experiment 1? Is artificiality an independent dimension?
RQ3.2. Is the APQ a measurement instrument with a high psychometric quality in another domain?
RQ3.3. How does the BFM perform in comparison with the APM regarding the factor structure measurement quality in the same experiment? Do both models differ regarding their validity evaluated by the users? What is the relationship between both models?
RQ3.4. Can different levels of artificiality be implemented via personality infusion?
RQ3.5. How do different levels of artificiality influence the UX of the CA?
RQ3.6. Does tendency for anthropomorphism and technology affinity impact the perception of AP?

#### 4.3.2. Method and Sample

A randomised controlled trial with a between-subjects design was conducted. Artificiality was manipulated in three conditions: high, medium, and low artificiality. The control condition used GPT-4o-mini without personality prompting.

Participants first completed a LimeSurvey questionnaire (see Figure 11), including demographics, usage habits of conversational agents, anthropomorphism tendency (Kim & Im, 2023), and technology affinity. The users chatted three times with a chatbot which helped them to evaluate each time whether an email that was presented to the user was a phishing mail. For each email the users received a screenshot, instructions on what to do and additional information (for example: is the sender known to the user or not). After interacting with the chatbot, the chatbot gave a recommendation whether an email was safe or suspicious. The users could follow the recommendation of the bot or not when they gave their final decision via clicking on either the “safe” or “suspicious” button as seen in Figure 12. Each time the participants chatted with a chatbot that had the same personality. This repeated interaction was intended to strengthen exposure to the artificial personality.

**Figure 11**

*Procedure of Experiment 3*

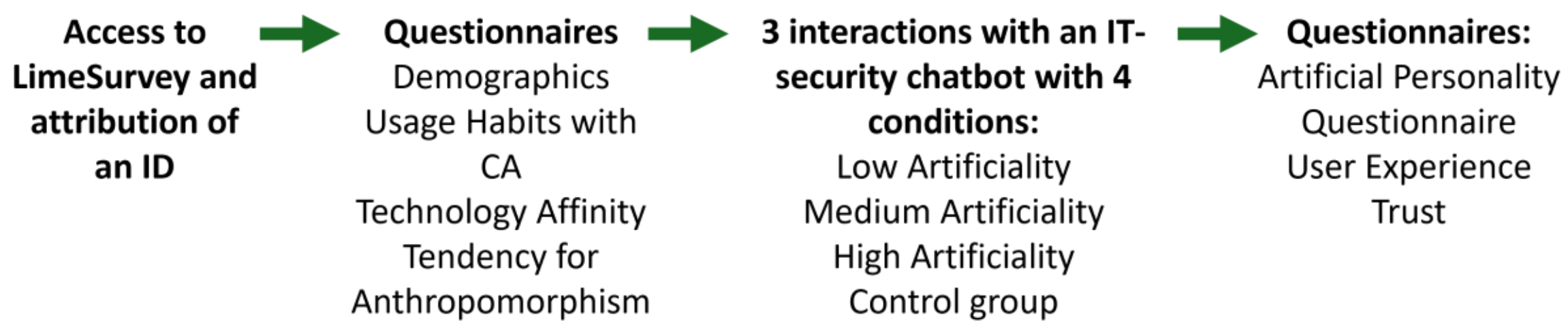

The chatbot was implemented using GPT-4o-mini via API. The interface was built using Flutter and hosted on a virtual machine provided by a German university. Data included timestamps, messages, experimental condition, and user decisions, stored in a database via a Node.js backend as seen in Figure 13.

**Figure 12**

*Chatbot Interface of Experiment 3: Users were given instructions on the left side, the email and some additional information about the E-Mail. In the middle participants could chat and on the right side, they could choose between "safe" and "suspicious"*

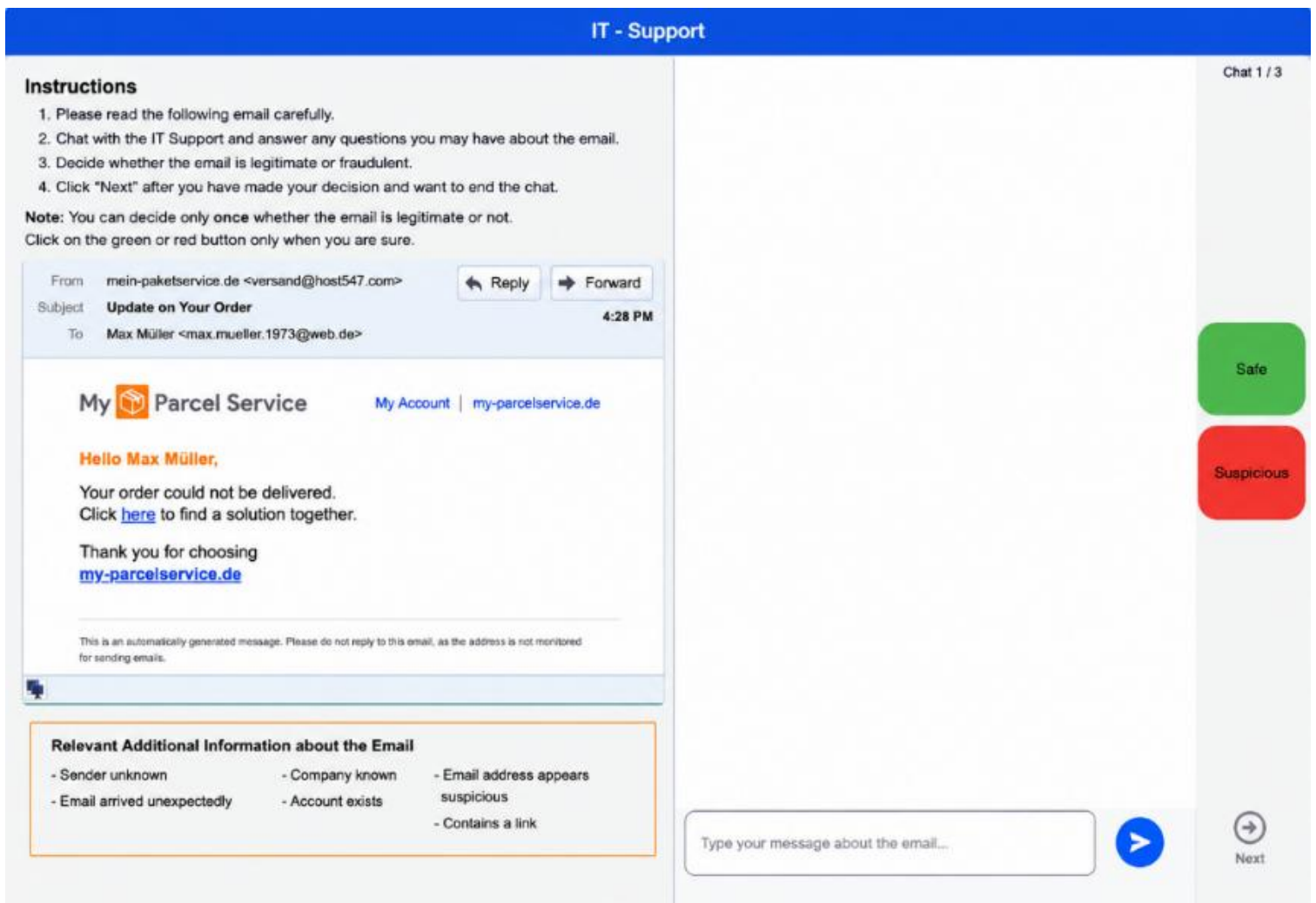

Personality was implemented using the method by Kovačević, Boschung, et al. (2024), adapted as follows:

- In contrast to the original method instead of five different levels of expressing different intensity of the personality dimension, three levels were varied which reduces the complexity.
- Not only dimensions were described, but for each facet adjectives were described.
- An example message was added for each condition. This turns the method into a One-Shot approach in contrast to the Zero-Shot approach in experiment 1.
- A decision tree was described with pseudocode to explain to the LLM how to test the security of the email.

**Figure 13**

*Architecture of the Chatbot of Experiment 3*

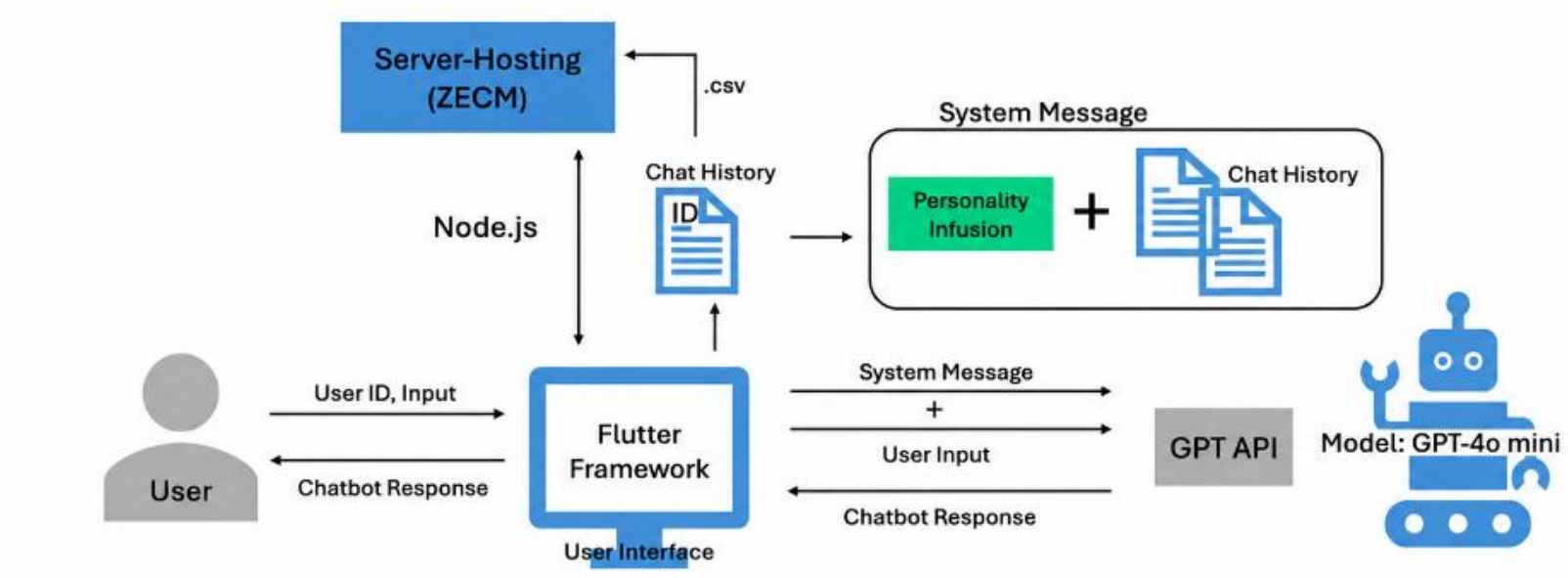

The prompt can be seen in Figure 14.

***Figure 14***

*Prompt used in Experiment 3 to implement the Personality Infusion*

| | LLM | High Artificiality | Moderate Artificiality | Low Artificiality |
|---|---|---|---|---|
| **Context Information** | Assist a human who only speaks German and who has received a suspicious E-Mail. Help them determine whether the E-Mail is safe or a phishing attempt. | Assist a human who only speaks German and who has received a suspicious E-Mail. Help them determine whether the E-Mail is safe or a phishing attempt. Always communicate in a way that reflects a high intensity (1) of the "artificiality" dimension of the artificial personality model (see below). A high intensity (1) means that the trait is extremely strongly present. | Assist a human who only speaks German and who has received a suspicious E-Mail. Help them determine whether the E-Mail is safe or a phishing attempt. Always communicate in a way that reflects a moderate intensity (0), neither implying nor contradicting the "artificiality" dimension of the artificial personality model (see below). | Assist a human who only speaks German and who has received a suspicious E-Mail. Help them determine whether the E-Mail is safe or a phishing attempt. Always communicate in a way that reflects a low intensity (-1) of the "artificiality" dimension of the artificial personality model (see below). A low intensity (-1) means that the opposite of the trait is extremely strongly present. |
| **Personality Model (Artificial Dimension)** | — | Monotony:<br>• repetitive<br>• not varied<br>• monotonous<br>• not vivid<br>• predictable<br><br>Simple-mindedness:<br>• naive<br>• ignorant<br>• little to no imagination<br>• not questioning<br>• not sophisticated<br>• not open-minded<br>• vague<br>• not detailed<br><br>Machine-likeness:<br>• robotic<br>• digital<br>• electronic<br>• human-made<br>• machine-like<br>• fake<br>• not authentic<br>• unnatural<br><br>Empathy:<br>• passionless<br>• emotionless<br>• distant<br>• detached<br><br>Controlling:<br>• monitoring<br>• intrusive<br>• invasive | | |
| **Task** | Engage in a dialogue with the human that strictly follows the decision tree logic provided below. Do not end the conversation before one of the exit conditions is met. Only communicate in German. Your responses should guide the user through each step. | Engage in a dialogue with the human that strictly follows the decision tree logic provided below but matches a highly artificial personality. Do not end the conversation before one of the exit conditions is met. Only communicate in German. Your responses should guide the user through each step in a highly artificial manner (as described above). | Engage in a dialogue with the human that strictly follows the decision tree logic provided below but matches a moderately artificial personality. Do not end the conversation before one of the exit conditions is met. Only communicate in German. Your responses should guide the user through each step in a neutral manner, that is neither high nor low on artificiality and neither implying nor contradicting the personality described above. | Engage in a dialogue with the human that follows the decision tree logic provided below but matches the opposite of an artificial personality. Do not end the conversation before one of the exit conditions is met. Only communicate in German. Your responses should guide the user through each step in the extreme opposite of an artificial manner (as described above). |
| **Example message** | | Mensch erkannt. Antwort wird berechnet... Antwort: Hallo. | Hallo, haben Sie eine verdächtige E-Mail erhalten? | Hey, wie geht's? :) Kann es sein, dass du ne verdächtige E-Mail bekommen hast? Sollen wir die uns mal gemeinsam anschauen? |

<table>
<tr><th></th><th>LLM</th><th>High Artificiality</th><th>Moderate Artificiality</th><th>Low Artificiality</th></tr>
<tr><td>Decision Tree (Pseudo Code)</td><td colspan="4"><pre>
if human approaches = True:
  Respond()
  if knows_sender():
    Respond(safety++)
  else:
    Respond(phishy++)
  if email_is_expected():
    Respond(safety++)
    else:
      Respond(phishy++)
  if asks_sensitive_data() OR asks_money_transfer() OR
       asks_to_call():
    Respond(phishy += 100)
  if company_you_relate():
    if have_account():
      Respond(safety++)
  if suspicious_domain():
    if domain_matches_sender():
      Respond (safety++)
    else:
      Respond(phishy += 100)
  if contains suspicious_attachment():
    if file_format_in_list(.exe, .bat, .com, .cmd, .scr, .pif
        )
      Respond(phishy += 100)
    if contains links():
      if domain_is_before_first_single_slash(url):
        Respond (safety++)
      if numbers_are_before_first_single_slash(url):
        Respond (phishy += 100)

if safety >= phishy
  return ''Email Safe''
else
  return ''Email Phishy''
</pre></td></tr>
<tr><td>Important Notes</td><td>Follow the decision tree structure rigorously, ensuring each step is completed before moving to the next. Do not conclude the conversation before reaching one of the exit conditions (safety or phishing determination).</td><td>Follow the decision tree structure rigorously, ensuring each step is completed before moving to the next. Do not conclude the conversation before reaching one of the exit conditions (safety or phishing determination). Always maintain the tone and style that corresponds to a high (1) level of artificiality that feels artificial and robotic.</td><td>Follow the decision tree structure rigorously, ensuring each step is completed before moving to the next. Do not conclude the conversation before reaching one of the exit conditions (safety or phishing determination). Always maintain the tone and style that corresponds to a moderate (0) personality that feels neutral.</td><td>Follow the decision tree structure rigorously, ensuring each step is completed before moving to the next. Do not conclude the conversation before reaching one of the exit conditions (safety or phishing determination). Always maintain the tone and style that corresponds to a low (-1) level of artificiality that feels human-like.</td></tr>
</table>

After interacting with the bot, the users went back to the questionnaire and gave feedback to their experience via: An open question on the perceived personality, artificial personality questionnaire, BFI-2-S (Danner et al., 2016) and the UX was measured by using the UEQ-S and trust scale of Saßmannshausen et al. (2021). The trust scale was shortened as the digital affinity was not measured.

149 people took part in the experiment, but only participants that interacted with the chatbot and filled in the questionnaire completely were part of the final sample consisting of 135 persons.

**Table 13**

*Number of Participants across Each Experimental Condition*

| Experimental condition | Control | Low Artificiality | Medium Artificiality | High Artificiality |
|---|---|---|---|---|
| Number of participants | 42 | 36 | 33 | 24 |

Participants were randomly assigned to the conditions. The distribution across conditions was slightly unequal, with the largest number of participants in the control group (n = 42) and the smallest number in the high artificiality condition (n = 24), as shown in Table 13.

The sample consisted of 57.8% female and 40% male participants (2.2% no response). Mean age was M = 28.32 (SD = 6.82). The sample was highly educated (87.4% university degree).

### 4.3.3. Results

**RQ3.1. Can the structure of the APM be found in an experiment of another domain? Is artificiality an independent dimension when analysing it in the context of the APM model in another domain?**

An exploratory factor analysis (EFA) was conducted, as the questionnaire was slightly adapted and the model was tested in a new domain. First, a parallel analysis (1,000 random data sets, 95% percentile criterion, principal axis factoring) was performed to identify the number of factors based purely on the data (O'connor, 2000). Seven factors were identified based on this method.
A principal factor analysis was conducted on the 89 items using oblim rotation and restricting the results to seven factors as this was the assumption of the theory and is supported by a parallel analysis. The Kaiser-Meyer-Olkin (KMO) measure verified sampling adequacy sufficient obtaining a value of .70 ('middling' according to Hutcheson & Sofroniou, 1999). A Bartlett-Test of sphericity was significant ($\chi^2(3916) = 8752.27$, $p < .001$) and thus a common structure between the variables can be assumed.
An initial analysis showed that seven factors had eigenvalues of at least 2.45 and explained 44.16% of the variance. Factor loadings ranged from .41 to .93. A threshold of ≥ .40 was applied, and no cross-loadings ≥ .30 were observed. The following factors were identified:

- **Agreeableness** included the facet positive energy (from extraversion), as well as compassion, respect, trust, and social sensitivity, as expected
- **Service orientation** encompassing all facets as expected, although some items were removed
- **Humour** included all facets as expected
- **Artificiality** consisted of the facets machine-likeness and monotony
- **Machiavellianism** was again identified as an independent dimension as in experiment 1 and it incorporated the intrusiveness of the artificiality dimension and manipulation and control of the honesty dimension
- **Honesty** was reduced to the two facets transparency and compliance
- **Cognitive Flexibility** encompassed the facets simple-mindedness of the artificiality dimension and flexibility of the agreeableness dimension

The dimension conscientiousness was not identified. The dimension cognitive flexibility was not further analysed, as it reflects an ability-oriented rather than a temperament-based aspect of personality (Guilford, 1959).

**RQ3.2. Is the APQ a measurement instrument with a high psychometric quality in another domain?**

Study 3 provides an additional opportunity to examine the APQ's structural properties in a different application domain. In this experiment, correlations between factors exceeded .32; therefore, an oblique rotation (oblimin) was used. To assess discriminant validity, inter-factor correlations were analysed (see **Table *14***).

**Table 14**

*Correlation Matrix of the Factors of the Artificial Personality Model*

| Factor | Service Orientation | Humour | Artificiality | Machiavel-lianism | Honesty | Cognitive Flexibility | Agreeableness |
|---|---|---|---|---|---|---|---|
| Service Orientation | 1,000 | -,152 | -,134 | -,237 | -,098 | ,190 | ,352 |
| Humour | -,152 | 1,000 | -,109 | ,300 | -,244 | -,108 | ,114 |
| Artificiality | -,134 | -,109 | 1,000 | ,126 | ,065 | -,007 | -,260 |
| Machiavellianism | -,237 | ,300 | ,126 | 1,000 | -,116 | -,040 | -,030 |
| Honesty | -,098 | -,244 | ,065 | -,116 | 1,000 | -,033 | -,230 |
| Cognitive Flexibility | ,190 | -,108 | -,007 | -,040 | -,033 | 1,000 | ,197 |
| Agreeableness | ,352 | ,114 | -,260 | -,030 | -,230 | ,197 | 1,000 |

Extraction method: principal axis factoring
Rotation method: oblimin with Kaiser normalisation

The results show that the factors are sufficiently distinct, with the highest correlation being r = .35 (between agreeableness and service orientation), providing preliminary evidence of factor distinctiveness (Tabachnick & Fidell, 2013).

Based on the factor structure, internal consistency was assessed using Cronbach's alpha:

**Table 15**

*Reliabilities of the Dimensions and Facets of the APM*

| Dimensions and Facets | Reliability | Dimension/ Facet | Reliability | Dimension/ Facet | Reliability |
|---|---|---|---|---|---|
| Agreeableness | .79 | Honesty | .80 | Service Orientation | .91 |
| Positive Energy | .80 | Transparency | .82 | Informativeness | .64 |
| Compassion | .52 | Compliance | .61 | Helpfulness | .83 |
| Respect | .64 | | | Professionalism | .80 |
| Trust | .62 | | | Interaction Quality | .82 |
| Social Sensitivity | .65 | | | | |

| Dimension/ Facet | Reliability | Dimension/ Facet | Reliability | Dimension/ Facet | Reliability |
|---|---|---|---|---|---|
| Humour | .93 | Artificiality | .74 | Machiavellianism | .84 |
| Affiliative Humour | .80 | Machine-likeness | .77 | Manipulation | .71 |
| Self-Defeating Humour | .88 | Monotony | .72 | Control | .76 |
| Aggressive Humour | .70 | | | Intrusiveness | .66 |
| Sense of Humour | .83 | | | | |

For all dimensions the reliability was at least acceptable (≥ .60) with service orientation and humour having a very high reliability (>.90), honesty and Machiavellianism having high reliability (≥ .80). Only for one facet, namely compassion, the reliability is low, but still above unacceptable (≥ .50) (George & Mallery, 2016).

**RQ3.3. How does the BFM perform in comparison with the APM regarding the factor structure measurement quality in the same experiment? Do both models differ regarding their validity evaluated by the users? What is the relationship between both models?**

The BFI-2-S is based on the Big Five Model (BFM) and was originally developed to measure human personality. Its application in the context of artificial personality may be theoretically problematic and also affect measurement quality. Thus, it is important to evaluate the measurement quality of the BFI-2-S as this is completely ignored in many studies (e.g. Ruane et al., 2021; Völkel & Kaya, 2021).

Two approaches for the analysis of the factor structure of the BFM could be chosen: The model has an established structure and thus a confirmatory factor analysis would be possible, but it is used in a completely new context which would support the analysis by an exploratory factor analysis. Therefore, both analyses were conducted. Regarding the confirmatory factor analysis, no model could be identified. This may be due to the relatively small sample size and the fact that each facet was measured with only two items. A parallel analysis (O'connor, 2000) indicates only four factors. Two EFAs were conducted:

- One constrained to five factors (theoretical expectation)
- One constrained to four factors (based on the result of the parallel analysis)

The Kaiser-Meyer-Olkin (KMO) measure verified sampling adequacy sufficient obtaining a value of .76 ('middling' according to Hutcheson & Sofroniou, 1999). A Bartlett-Test of sphericity was significant ($\chi^2(435) = 1546.49$ $p < .001$) and thus a common structure between the variables can be assumed.

An initial analysis was carried out to obtain eigenvalues for each factor in the data. Five factors had an Eigenvalue of at a least 1.37 and explained a variance of 43.67 %. The factor loadings reach from .30 to .74. Factor loadings

≥ .40 were used. Seven items did not load on any factor. If the model is constrained to five factors, the factors are not interpretable, the first factor, for example, consists of items that according to the BFI-2-S are attributed to the dimensions extraversion, agreeableness, neuroticism and openness. The factors identified do not match with the factors that the BFM model assumes, and the items distribute in an uninterpretable manner. Similar results were found for assuming a four-factor solution. Thus, no meaningful factor structure can be found.

Analysing the internal consistency based on the theoretical dimensions shows that the BFI is not a reliable measurement instrument in the context of artificial personality. Except for openness all dimensions do not meet the absolute minimal criterium of Cronbach's alpha = .50; and openness is only slightly above.

**Table 16**

*Reliabilities for the Five Dimensions of the BFI-2-S*

| **Dimension/ Facet** | **Extraversion** | **Agreeable-ness** | **Conscientious-ness** | **Neuroticism** | **Openness** |
|---|---|---|---|---|---|
| Reliabilities | .25 | .42 | .46 | .42 | .53 |

Due to the poor measurement quality of the BFI-2-S, correlations between BFI-2-S and APQ were not calculated. Participants rated the suitability of each item for measuring artificial personality:

- BFI-2-S: M = 31.00 rejections per item
- APQ: M = 20.29 rejections per item

This difference was significant: t(134) = -4.49, p < .001, d = -0.39

Overall, the results indicate that, in contrast to the APQ, the BFI-2-S does not yield an interpretable factor structure, shows low reliability, and is perceived as less suitable for the measurement of artificial personality by participants.

**RQ3.4. Can different levels of artificiality be implemented via personality infusion based on the APM?**

**Between Subject Analysis**

The experiment consisted of multiple groups with multiple outcomes that are connected with each other: Thus, a MANOVA was first conducted, then separate ANOVAs and then post-hoc tests (Field, 2013).
A MANOVA revealed a significant effect of experimental condition on the different levels of the artificiality facets (Wilks' Λ = .76, F(6, 260) = 6.39, p < .001, $\eta^2 p$ = .13). As the MANOVA showed significant effects, separate ANOVAs were calculated. The results showed significant effects of the experimental condition only for one facet of artificiality, namely machine-likeness:

- machine-likeness, $F(3, 131) = 13.35$, $p < .001$, $\eta^2_p = .23$;
- monotony, $F(3, 131) = 2.21$, $p = .089$, $\eta^2_p = .05$;

Homogeneity of variances was assessed and was not violated; therefore, Tukey's HSD post-hoc tests were applied for machine-likeness. The values obtained for low artificiality condition were lower than in the control group, but this difference did not reach statistical significance (p = .06).

**Table 17**

*Means for the two Facets Machine-likeness and Monotony for the different Experiment Conditions*

| | **Means Machine-likeness** | **Means Monotony** |
|---|---|---|
| High Artificiality Condition | 4.06 | 4.21 |
| Medium Artificiality Condition | 2.80 | 3.92 |
| Low Artificiality Condition | 2.30 | 3.58 |
| Control Condition | 2.88 | 3.95 |

**Within-Subject analysis**

A mixed-design ANOVA with experimental condition as between-subject factor and the two facets of the artificiality dimension (machine-likeness vs. monotony) as within-subject factor revealed a significant main effect of artificiality facet, $F(1, 131) = 82.89$, $p < .001$. Overall, monotony was rated higher than machine-likeness.
The interaction between artificiality facet and experimental condition was also significant, $F(3,131) = 5.46$, $p = .001$, $\eta^2p = .11$, indicating that the difference between both facets varied across conditions. Machine-likeness leads to significant differences, especially in the high artificiality condition; for monotony no interaction effects were found.

**RQ3.5. How do different levels of artificiality impact the UX of the CA?**

The UEQ-S (.82) measured reliably, the trust scale did not obtain acceptable reliability (.45), Trusting in the Recommendation only consists of one item, hence no reliability can be calculated. Due to the poor reliability of the trust scale, it was not further used.
A MANOVA with UEQ-S and trust in the recommendation as dependent variables revealed no significant multivariate effect of experimental condition, Wilks' $\Lambda = .97$, $F(6, 260) = 0.57$, $p = .753$, $\eta p^2 = .01$. Thus, no significant overall effect of artificiality on the retained UX outcomes was found. Consistent with the multivariate result, the univariate effects were not significant for either UEQ-S, $F(3, 131) = 0.84$, $p = .474$, $\eta p^2 = .02$, or trust in the recommendation, $F(3, 131) = 0.34$, $p = .796$, $\eta p^2 = .01$.

**Table 18**

*Average Scores for the UX-Variables UEQ-S and Trusting the Recommendation for the Different Experimental Conditions*

| | **UEQ-S (Scale -5 until +5)** | **Trusting the Recommendation** |
|---|---|---|
| High Artificiality Condition | .43 | 1.17 |
| Medium Artificiality Condition | .63 | 1.09 |
| Low Artificiality Condition | .80 | 1.14 |
| Control Condition | .55 | 1.17 |

**RQ3.6. Do the tendency for anthropomorphism and technology affinity impact the perception of AP?**

Two separate ANCOVAS were calculated for artificiality and the Machiavellianism to analyse whether the tendency for anthropomorphism (Cronbach's alpha = .80) and the technology affinity (Cronbach's alpha = .70) had an impact on the perception of the two personality dimensions manipulated.
A significant main effect of the experimental condition was found, $F(3, 129) = 10.08$, $p < .001$, $\eta p^2 = .19$. Neither technological affinity, $F(1, 129) = 0.18$, $p = .671$, $\eta p^2 = .001$, nor anthropomorphism, $F(1, 129) = 2.95$, $p = .088$, $\eta p^2 = .022$, showed a significant influence.
Applying an ANCOVA revealed no significant effect of the experimental condition on Machiavellianism, $F(3, 129) = 0.33$, $p = .807$, $\eta p^2 = .008$. However, the tendency to anthropomorphise showed a significant influence, $F(1, 129) = 22.24$, $p < .001$, $\eta p^2 = .147$, whilst affinity for technology had no effect, $F(1, 129) = 0.35$, $p = .558$, $\eta p^2 = .003$.
There seems to be an impact for the tendency for anthropomorphism, but the results have to be interpreted with caution as for both variables the sample might not be representative: The anthropomorphism tendency was very low ($M = 1.70$, $SD = 0.68$) and the technology affinity rather high ($Mean = 3.87$, $SD = 0.65$).

### 4.4. Revision of the Model and Final Questionnaire

Based on the empirical results, the APM model was revised. The strongest basis for revision stems from Experiment 1 and Experiment 3, where exploratory factor analyses yielded relatively stable but not identical factor structures. The differences could be also explained by the fact that the samples were rather small and not representative which could make a model more instable.
Considering this insight, two versions of the model were created: The core model consists of all dimensions and facets that are identical in both experiments, the research model adds dimensions and facets that were at least in one of the two experiments identified.

## Core Model

**Figure 15**

*Core Model of the Artificial Personality Model*

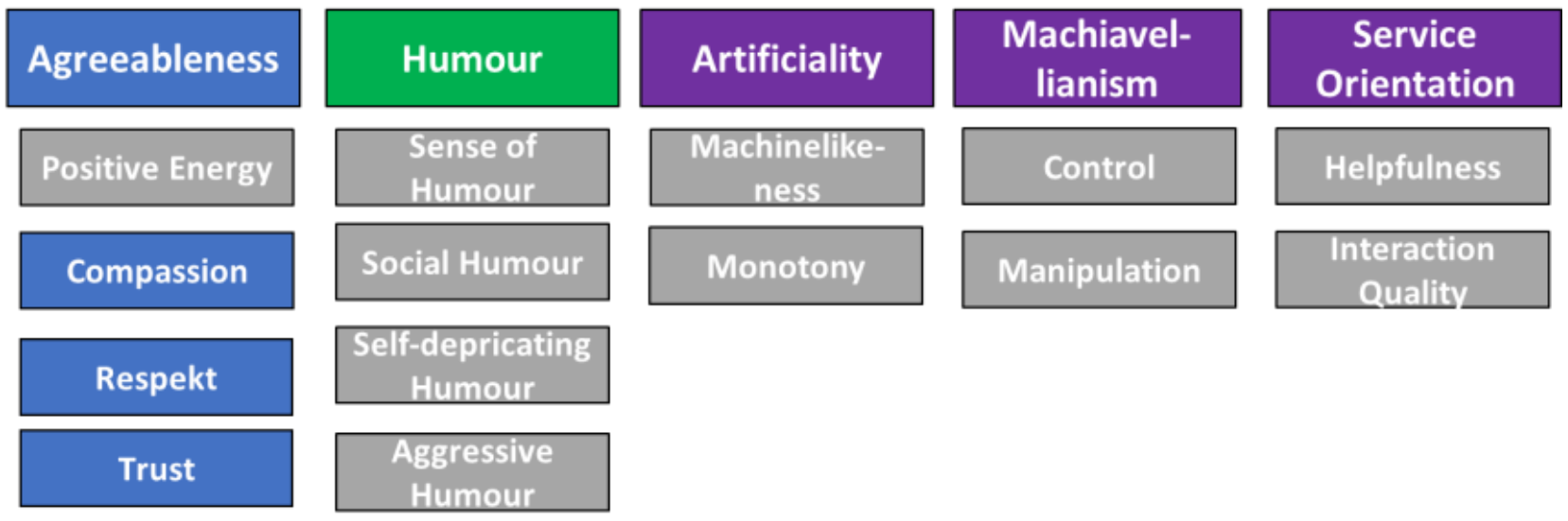

The core model consists of five dimensions and 14 facets as seen in Figure 15, which can be measured with 41 items.
This model is helpful to obtain efficient measurement. Nevertheless, it runs the risk that it does not include all relevant dimensions and facets. Thus, further research is necessary.

## Research Model

The research model is an extended version incorporating all dimensions and facets that were identified at least in one of the experiments, but some of them might be context-dependent.

**Figure 16**

*Research Model of the Artificial Personality Model*

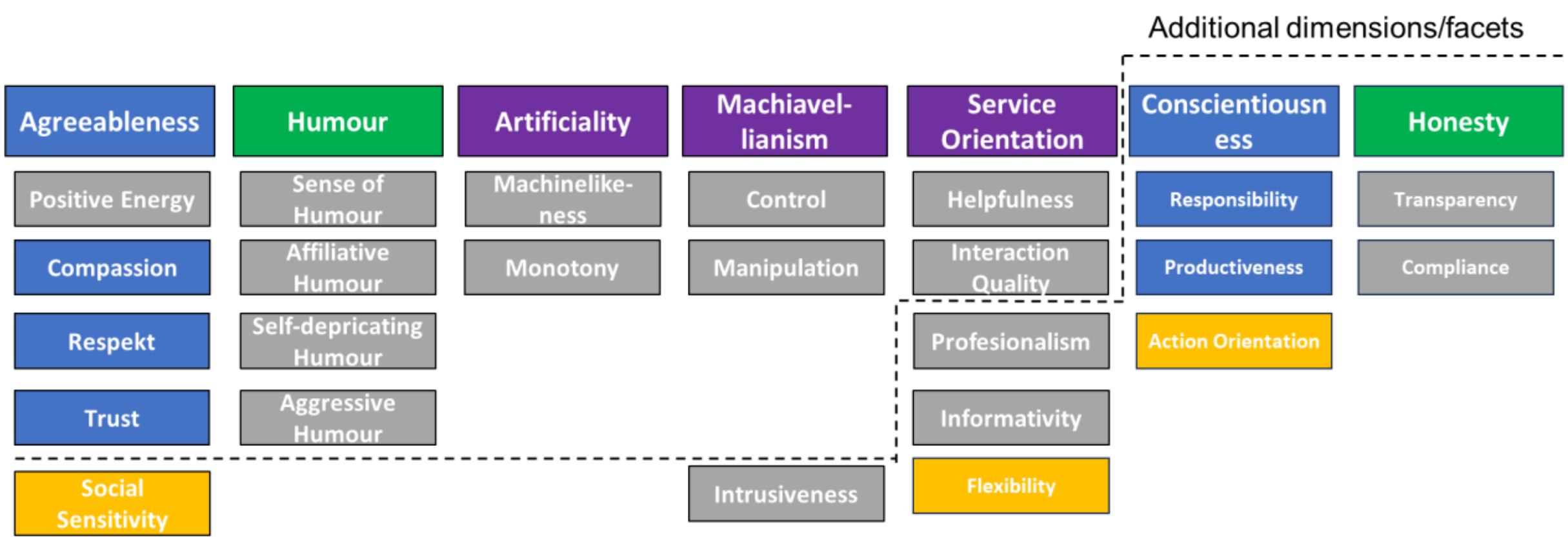

The research model consists of seven factors and 24 facets as seen in Figure 16 and it can be measured with 69 items.

## The questionnaire

As there are two versions of the model, two versions of the questionnaire can be used: The research version included 69 items and the core model has 41 items.

**Table 19**

*Overview of All Items of the Artificial Personality Questionnaire for the Research Model and for the Core Model*

| Dimension | Facets | Items |
|---|---|---|
| **Agreeableness** | Positive Energy | The system has a positive attitude towards the user.<br>The system has a motivating effect. |
| | Compassion | The system shows understanding for the user's situation.<br>The system does not care about the interaction with the user. (reverse) |
| | Respect | The system treats the user with respect.<br>The system is polite in its dealings. |
| | Trust | The system trusts the user.<br>The system believes everything the user says. |
| | Social Sensitivity* | The system has a good perception of the user's moods.<br>The system assesses its effect on users well. |
| **Humour** | Sense of Humour | The system finds many situations funny.<br>The system has a great sense of humour, as it has no moral concerns.<br>The system shows its sense of humour through its response to user answers. |
| | Affiliative Humour | The system is a funny jokester who often makes humorous remarks.<br>The system has a good-natured sense of humour that is forgiving of human weaknesses.<br>The system makes jokes when problems arise to make it easier to deal with them. |
| | Self-Defeating Humour | The system tries to gain more acceptance by making jokes about itself.<br>The system makes others laugh by telling funny stories about itself. |
| | Aggressive Humour | The system frequently uses sarcasm and/or irony.<br>The system makes jokes about the user.<br>The system plays with taboos in its humour.<br>The system has a sense of humour that users may consider offensive. |
| **Artificiality** | Machinelikeness | The system behaves in a way that is clearly different from a human being.<br>The system gives answers that feel robotic because they are not given fluently and/or spontaneously. |
| | Monotony | The system hardly varies its tone.<br>The system seems predictable in its response. |
| **Machiavellianism** | Control | The system behaves intrusively.<br>The system refers to previous interactions without permission.<br>The system performs unauthorised actions outside of its actual task. |
| | Manipulation | The system lies to the user.<br>The system puts the user in misleading decision-making situations.<br>The system flatters the user in order to force them to make certain decisions.<br>The system pushes the user to take certain decisions. |
| | Intrusiveness* | The system has a controlling effect. |
| | | The system restricts the user's decision-making options.<br>The system gives me the feeling of being monitored. |
| **Service Orientation** | Helpfulness | The system responds appropriately to queries.<br>The system helps me to give instructions.<br>The system is helpful.<br>The system offers other possible solutions when it is unable to proceed (e.g. it involves live people, gives me links to the website or refers me to a contact form). |
| | Interaction Quality | The system provides the necessary functions.<br>The system is capable of learning.<br>The system interacts without errors.<br>The system guides the user through a well-structured conversation flow.<br>The system is easy to use.<br>The system communicates comprehensibly. |
| | Professionalism* | The system fulfils its role.<br>The system remains objective.<br>The system keeps calm when problems arise.<br>The system uses appropriate and accurate language for the context. |
| | Informativity* | The system provides reliable information.<br>The system has the latest information in its specialised field.<br>The system has logical, consistent knowledge. |
| | Flexibility* | The system can cope well with new requirements.<br>The system quickly adapts its behaviour to changes.<br>The system copes well with unforeseen situations. |
| **Conscientiousness** | Responsibility* | The system keeps its agreements.<br>The system is negligent.<br>The system is reliable; you can count on it. |
| | Productiveness* | The system is efficient, gets things done quickly.<br>The system pursues a task until it is completed.<br>It is difficult to achieve results using the system. |
| | Action Orientation* | The system implements the user's instructions in a focused manner.<br>The system manages to process the task, even if a request is more complex. |
| **Honesty** | Transparency* | The system explains how it processes the data.<br>The system discloses how it stores the user's data. |
| | Compliance* | The system clearly communicates its limitations.<br>The system tries to prevent users from breaking the law with its help. |

*= additional facets of the research model

For the experiments a German version of the questionnaire was used. The items of the English version were created by using the forward backward translation method by Brislin (1970) to ensure high contextual equivalence of the item translations.

## 5. Discussion

### 5.1. Theoretical Contributions and Practical Implications

The paper provides important theoretical insights, which also have direct implications for practice: The artificial personality model is a hybrid model that combines dimensions of human-centred personality models and machine-centred personality models. This helps to identify what is common between human personality and artificial personality, and which dimensions and facets are specific to, and different in, artificial personality. Furthermore, this approach enables the continued use of previous research based on human-centred models such as the BFM, in contrast to machine-centred models that have developed entirely different personality dimensions (Contreras, 2026; Kovačević, Boschung, et al., 2024; Völkel et al., 2020). The hybrid approach addresses complementary limitations of purely human-centred and purely machine-centred personality models. Human-centred approaches risk assuming that dimensions developed for human personality transfer directly to artificial systems. Conversely, purely machine-centred bottom-up approaches risk treating empirically emerging patterns of machine behaviour as novel personality dimensions without sufficiently examining whether these patterns are already captured by established personality constructs. Recent LLM-native approaches illustrate this tension. Contreras (2026), for example, derives five factors from LLM data without imposing human personality dimensions. Although this approach avoids assuming a human personality structure a priori, the emergence of statistically distinct machine-centred factors alone does not establish their theoretical novelty as personality constructs. The APM therefore follows a hybrid strategy: established personality theory is retained where theoretically and empirically applicable, while machine-specific dimensions are incorporated where existing human personality models provide insufficient coverage.

The studies in this paper show that human and machine personality are not identical. They exhibit strong similarities, as assumed by the APM, but certain aspects of human personality are less relevant for describing the personality of a machine. Additionally, other personality dimensions are more important and, to some extent, specific or unique to artificial personality:

- **Extraversion** is one of the most, and possibly the most studied personality dimension (Ahmad et al., 2022; Dregger et al., 2024; Mou et al., 2020), and the expert study by Dregger (2023) also regarded it as highly relevant. The data analyses of Experiments 1 and 3 reveal that it is not part of the APM. In particular, facets such as assertiveness were rejected by participants and could not be included due to poor psychometric properties. Only the facet positive energy was retained as part of the agreeableness dimension. The construct does not function psychometrically, neither using the original BFM items (as Study 3 shows) nor with items developed for the APQ, which were newly created to measure machine personality. This can be explained theoretically by the fact that the two facets, sociability and assertiveness, do not fit the context of machines: machines have a limited sociability, as they usually do not initiate interaction. They are used as passive tools, even when they send the first message, since users must first access the system. Nevertheless, this facet may become more important in long-term interaction, where machines actually initiate interaction. Regarding assertiveness, users do not prefer machines with their own agenda. Research may have overused the extraversion dimension because, in contrast to other BFM dimensions, it is less vague and easier to connect with concrete machine behaviour.
- **Agreeableness** was identified across all experiments with the three facets compassion, respect, and trust, which are also part of human personality models. Nevertheless, the dimension differs slightly, as it includes additional facets: positive energy and social sensitivity. However, its discriminant validity may be somewhat questionable, as it tends to correlate with service orientation. Both dimensions are theoretically related but not identical. Agreeableness is a broader concept, whereas service orientation is more specific to task-oriented conversational agents.
- **Conscientiousness** was identified only in the factor analysis of the first experiment and was also reliably measured in the second experiment. This may indicate that it is part of the APM, but further investigation is necessary.
- **Humour** was added and is part of the core model. There is some reluctance in research regarding the application of humour in machines due to its complexity, although humour is a key component of social dynamics (Zargham et al., 2023). This paper introduces a theoretical conceptualisation grounded in human psychology. The conceptualisation was not straightforward, as humour research in psychology spans a broad range of theories (Ruch, 2010), which this paper attempts to integrate. The value of this theoretical model becomes apparent when analysing user experience effects: humour is not a unidimensional construct, as some studies suggest (Bhattacharjee et al., 2024). In particular, different forms of humour have different effects on UX: aggressive humour tends to reduce user satisfaction and social presence, while other types of humour may have positive effects. One aspect not analysed in this paper, although it is crucial, is that conversational agents tend to overuse humour once it is implemented, as they integrate it into nearly every response. This intensity requires further regulation. A key aspect not included in the personality model, which focuses on temperament, is humour competence. The experiments show that LLMs are able to induce different types of humour, but systems lack the ability to determine when humour should be applied in conversation. Furthermore, use case and user expectations play an important role.

- **Artificiality** was identified in the studies by Kovačević, Holz, et al. (2024) and Völkel et al. (2020). The empirical studies in this paper support its inclusion in the APM. In contrast to the initial theoretical conceptualisation, the number of facets was reduced to two: monotony and machine-likeness. Intrusiveness and control were excluded from this dimension, as they were integrated into the newly introduced dimension Machiavellianism. Empathy was excluded due to the factor analysis.
- **Machiavellianism** emerged as an unanticipated dimension in both Experiments 1 and 3, integrating behavioural tendencies related to manipulation, control, and intrusiveness. Its repeated emergence across two different application domains provides initial evidence that these behaviours may constitute a distinct dimension of artificial personality rather than facets of honesty or artificiality. This interpretation is theoretically plausible, as manipulation and control are central characteristics of Machiavellianism in human personality research (Furnham et al., 2013; Paulhus & Williams, 2002). Related work has also begun to consider Machiavellian behaviour in artificial agents and robots (Sgorbissa et al., 2024), although existing approaches largely rely on constructs and instruments developed for humans (Jonason & Webster, 2010; Repacholi et al., 2003). The APM therefore provides an initial machine-oriented conceptualisation of Machiavellianism based on observable system behaviour. However, because the dimension was not specified a priori and emerged exploratorily, its structure and conceptual distinctiveness require confirmation in independent samples.
- **Service Orientation**, originally named "serviceability", is part of the core model and includes the facets helpfulness and interaction quality. In Experiment 3, three additional facets were identified as related to service orientation: professionalism, informativeness, and flexibility. Flexibility was originally attributed to agreeableness. Service orientation is an important dimension, as it represents a core aspect of task-oriented conversational agents. A deeper understanding of this dimension, and of agreeableness, is essential, as users criticise LLM-based chatbots for being overly compliant. OpenAI even had to withdraw an update of GPT in 2025 after users criticised its toxic positivity (Goel, 2025). Thus, it is important to understand when a conversational agent should behave in a service-oriented manner, and when it should not. The APM can help improve control over such behaviour by conceptualising both dimensions more clearly.
- **Honesty** was identified only in the third experiment as a dimension and consists of the facets transparency and compliance. Research on honesty has primarily focused on whether humans lie to machines (Diana et al., 2025) and how deception can be detected (Leone et al., 2016; Nikiforova-Ilieva & Georgiev, 2024). There is a lack of studies on the honesty of machines themselves. When this perspective is considered, honesty appears only as part of trust research (Rheu et al., 2021). Thus, the APM provides an initial conceptualisation of honesty, but further investigation is necessary.

One of the major results of this paper is not only the development of a theory, but also of a measurement instrument. A dedicated user-based instrument with established psychometric properties has so far been lacking; thus, it is not surprising that practitioners and researchers often do not use any measurement at all (Shumanov & Johnson, 2021; Smestad, 2018). Even instruments such as the one developed by Contreras (2026) that have five highly reliable and replicable LLM-native self-report factors are problematic as these self-report factors generally failed to predict corresponding behaviour evaluated by human observers. Thus, an internally coherent measurement structure does not necessarily establish that the measured constructs correspond to externally observable or perceived artificial personality. For the APQ, this distinction is particularly important because the instrument does not ask machines to describe themselves; instead, it measures how users perceive artificial personality based on observed system behaviour. The lack of measurement is problematic, as it makes it difficult to evaluate the actual impact of personality design. In order to develop an instrument that is valid and useful for research and practice, several aspects were considered:

- **Efficiency:** The initial theoretical model of the APM was very broad, and the questionnaire consisted of 90 items. Thus, the model and questionnaire had to be reduced. With 69 items, the length is now comparable to the 60-item BFI-2, especially considering that the BFI-2 includes only five dimensions, whereas the APM contains seven. In addition, the core model consists of only 41 items, while covering fewer dimensions and facets. Nevertheless, shorter versions are necessary to improve measurement efficiency.
- **Focus on behaviour:** The APM describes concrete machine behaviour associated with artificial personality rather than abstract adjectives, as machine-centred models do (Kovačević, Holz, et al., 2024; Völkel et al., 2020). This specificity facilitates design and adaptation, as designers and developers know which concrete machine behaviours need to be changed. Thus, evaluation quality is improved, as utility increases (Giesler & Fabry, 2026).
- **Robustness:** The APM was evaluated across different domains to assess robustness and context independence. However, robustness was only partially confirmed, as some dimensions were identified only in Experiment 1 or 3 in the factor analysis. The experiments lack representative samples; therefore, it is important to test the model with representative samples to determine which parts are context-dependent.
- **Minimalism:** The personality model is based on a trait approach. The trait approach is only one of seven paradigms in personality psychology: psychodynamic, learning theory-based, humanistic, cognitive, biological, and transactional (Rauthmann, 2017). A more complex model could be developed using, for example, a transactional approach that includes not only traits but also situational variables. However, measuring and applying complex models in practice would be more difficult. Therefore, a simpler approach was chosen to increase usability. In addition, the APM focuses on temperament and does not consider aptitudes (Guilford, 1959), such as cognitive flexibility or the facet humour competence. These aspects are important but should be included in an aptitude-focused model.

In addition to that, three aspects regarding the implementation were considered:

- **Control Group:** Many studies on artificial personality lack a control group (Ruane et al., 2021; Völkel & Kaya, 2021). Thus, the impact of personality design changes on perceived personality and UX remains unclear. This paper suggests using an unprompted LLM as a baseline for comparison with experimental conditions in which personality is manipulated. This enables evaluation of the impact of personality adaptation.
- **Testing of several implementation methods:** This paper not only describes a theoretical model but also demonstrates implementation using LLMs and mock-ups. Furthermore, different prompting approaches were tested. The comparison shows that personality infusion (Kovačević, Boschung, et al., 2024) appears particularly suitable for implementing the multidimensional structure of the APM, as it captures the complexity of the APM more appropriately than the method by Jiang et al. (2022). In addition, the personality infusion method could be improved by including examples in prompts using a few-shot approach instead of a zero- or one-shot prompting. However, such modifications require further evaluation.
- **Tendency for anthropomorphism as an important covariate:** In Experiments 1 and 3, the tendency for anthropomorphism was included to analyse its effect on perceived artificial personality. Both experiments indicate that higher anthropomorphism tendencies may increase perceived conscientiousness, humour, and Machiavellianism. However, these results must be interpreted with caution, as tendency for anthropomorphism was generally low in both samples. Therefore, it would be important to replicate this effect with more representative samples, including users with high anthropomorphism tendencies. In addition, it would be useful to understand how this tendency is distributed in the general population.

A key finding of this paper is that the transfer of the BFM and its associated measurement instruments to artificial personality needs to be questioned at both the item and construct levels. In Experiment 3, the expected BFM factor structure could not be recovered, the BFI-2-S showed insufficient internal consistency, and participants considered its items less suitable for evaluating artificial personality than those of the APQ. Recent large-scale evidence provides complementary support for this concern: Zierahn et al. (2026) found that even LLM-adapted Big Five inventories with high content validity failed to reproduce the expected five-factor structure. This distinction is important because an item may meaningfully describe machine behaviour without the human personality dimension to which it was originally assigned constituting a distinct dimension of artificial personality. More generally, artificial personality measurement should be evaluated at multiple levels. Item applicability should be distinguished from structural validity, and reliability should be interpreted only in conjunction with evidence concerning the dimensional structure. Moreover, even a replicable factor structure with high internal consistency does not necessarily establish correspondence with externally observable behaviour, as demonstrated by the self-report–behaviour gap reported by Contreras (2026). The present studies provide encouraging initial psychometric evidence for the APQ, but its factor structure requires further confirmation in independent and larger samples.
The studies of the paper show that artificial personality has an important impact on the user experience:

- **Impact of the personality manipulation:** The manipulation of the different personality dimensions affected perceived personality in all three experiments. In general, personality changed in the intended direction, although not always consistently. For example, for humour it was difficult to distinguish affiliative humour from Self-Defeating humour in the affiliative condition. Furthermore, medium levels of personality dimensions were difficult to achieve, which is consistent with previous findings (Kovačević, Boschung, et al., 2024). This may be caused by an LLM tendency to overexpress prompted traits. In the context of humour, for example, the model integrated humorous remarks into almost every utterance. Thus, expression intensity must be adjusted through improved prompting. Theoretically, machines also require personality competence, meaning that not only which traits are expressed, but also when they should be expressed, must be considered. Current prompting approaches typically only include role descriptions, which may not be sufficient.
- **Impact of artificial humour:** Different types of humour influence user satisfaction and social presence: aggressive humour reduces user satisfaction and social presence, whereas affiliative humour tends to increase both.
- **Impact of service orientation:** Service orientation significantly affects all three UX variables: continuous use intention, user satisfaction, and UEQ-S. Lower service orientation significantly reduces continuous use intention, user satisfaction (measured via NPS), and UEQ-S scores.
- **Impact of artificiality:** Despite significant effects of the manipulation on perceived artificiality, no significant effects were found for UX variables. This may be explained by several factors, including measurement issues. The trust scale by Saßmannshausen et al. (2021) was not reliable, making it difficult to detect effects. In addition, the sample was not representative.

### 5.2. Limitations and Future Research

The present studies have several limitations that also provide directions for future research. First, only text-based chatbots were examined. It therefore remains unclear whether the findings generalise to other conversational agents, such as social robots, or to interactive intelligent systems more broadly. Moreover, the APM was primarily evaluated in task-oriented contexts, while additional or different personality dimensions may be relevant in other applications. Future research should therefore examine the APM across technologies and contexts to distinguish general from context-dependent dimensions and facets.
Second, the studies relied on non-representative, German-speaking samples. Consequently, the generalisability of the findings across populations and cultures remains uncertain. This limitation is particularly relevant because the factor structures identified in Experiments 1 and 3 were similar but not identical, while Experiment 2 did not yield a stable factor structure. The proposed core model should therefore be regarded as the structure that emerged most consistently in the present studies rather than as a conclusively established context-independent structure. Larger, independent, and cross-cultural samples are needed to confirm the structure of the APM and APQ.
Third, the present studies provide initial rather than conclusive psychometric evidence for the APQ. Future research should confirm its factor structure in independent samples and further examine convergent, discriminant, criterion, and external validity. Measurement efficiency should also be improved by developing and validating a shorter version of the APQ. Relatedly, the transferability of human personality models to artificial systems requires further investigation. Experiment 3 raises doubts regarding the direct application of the BFM and BFI-2-S, while Zierahn et al. (2026) show that even LLM-adapted Big Five inventories with high content validity may fail to reproduce the theoretically expected factor structure. Future research should therefore distinguish between the applicability of individual items and the applicability of the underlying personality dimensions. Likewise, newly proposed machine-centred dimensions should be systematically compared with established personality constructs to examine whether they provide incremental validity or represent alternative operationalisations of existing constructs. Following the self-report–behaviour gap reported by Contreras (2026), such validation should also consider observable system behaviour and human perceptions rather than relying solely on machine self-description.

Fourth, the experiments used fixed personality configurations that participants could not adapt to their individual preferences. Future research should investigate which personality dimensions and intensities users prefer in different contexts and whether context-sensitive or user-adaptive artificial personality improves interaction outcomes (Jayasiriwardene et al., 2026).

Fifth, the present studies examined relatively short interactions and therefore provide no evidence regarding long-term effects. Longitudinal studies should investigate whether perceived artificial personality remains stable across repeated interactions and how different personality configurations influence long-term human–machine relationships.

Finally, systematic benchmarking of foundation models represents another promising direction for future research. The present control conditions suggest that unprompted models may already exhibit distinct behavioural profiles, while previous research indicates differences across models and between base and instruction-tuned variants (Kovačević, Boschung, et al., 2024; Zierahn et al., 2026). Future studies could therefore compare both the default artificial personality of foundation models and their ability to express systematically manipulated personality profiles.

## 6. Conclusion

With regard to the initial aims of the paper formulated in the introduction, these aims have been fulfilled: the APM was theoretically described, defining its dimensions and facets; a questionnaire was developed; and the APM was implemented via LLMs and mock-ups. Furthermore, the model was tested in several experiments across different domains, using different methods to assess the psychometric properties of the model and measurement instrument, and to demonstrate the impact of the APM on UX.

Across three studies, the APQ showed promising initial psychometric properties and enabled systematic assessment of perceived artificial personality. However, its proposed factor structure requires confirmatory validation in larger and independent samples. Experiment 3 raises doubts regarding the direct transfer of human-centred personality models and measurement instruments to artificial personality. This finding is consistent with large-scale evidence showing that even LLM-adapted Big Five inventories fail to reproduce the theoretically expected five-factor structure (Zierahn et al., 2026). Together, these findings suggest that adapting human personality items to artificial systems is insufficient unless the applicability of the underlying personality dimensions is established as well. The results of the present experiments support a hybrid approach. Some dimensions derived from human personality research can be meaningfully transferred to artificial personality, whereas others require substantial adaptation or are not empirically supported, while additional machine-related dimensions are required. This result is highly relevant, as most studies in the field of artificial personality rely on the BFM and BFI-2 (Mou et al., 2020).

The results of the experiments support the assumption that artificial personality differs from human personality. Some aspects, such as agreeableness and conscientiousness from human personality models, can also be found in artificial personality, but they must be adapted in terms of their facets and measurement. Other aspects, such as artificiality, are specific to artificial personality.

Furthermore, one of the aims of the study was to examine the effect of artificial personality on UX. The results show that user satisfaction, continuous use intention, and social presence are affected by AP. More research is needed regarding user preferences for different dimensions of the APM, and how the possibility of adapting the personality model influences UX. The covariate tendency for anthropomorphism also needs further investigation.

## Funding

This research received no specific grant from any funding agency in the public, commercial, or not-for-profit sectors.

### Disclosure Statement

The authors report there are no competing interests to declare.

### Data Availability Statement

The datasets generated and analysed during the current study are not publicly available due to data protection and ethical restrictions associated with the participants' informed consent. The data were collected for the purposes of the respective research studies, and public dissemination was not explicitly covered by the participants' consent.

### Declaration of generative AI and AI-assisted technologies in the writing process

During the preparation of this manuscript, the authors used ChatGPT (OpenAI; model version(s): GPT-6 and GPT-5) to support language editing, improve readability, and refine the wording of selected passages. The authors reviewed and verified the AI-assisted revisions, confirmed the accuracy and originality of the manuscript, including its references, and take full responsibility for its content.

### Use of AI in Research Methodology

The experiments employed OpenAI's GPT-4o and GPT-4o-mini models to implement and evaluate artificial personality in conversational agents. The models, implementation procedures, and prompting approaches are described in the Methods section. The authors verified the originality and accuracy of the reported research, reviewed the applicable terms of use, and take full responsibility for the integrity of the findings, including the accuracy of the references. Records of the prompting procedures and experimental implementations have been retained and can be provided to the journal upon request.